\documentclass[11pt,titlepage]{article}
\usepackage{latexsym}
\usepackage{amsfonts}
\usepackage{amsmath}

\usepackage{amssymb}
\usepackage[margin=1in]{geometry}
\usepackage{graphicx}
\usepackage{tikz}
\usetikzlibrary{arrows, decorations.markings}
\usetikzlibrary{fit}					
\usetikzlibrary{backgrounds}

\newcommand\blackslug{\hbox{\hskip 1pt \vrule width 4pt height 8pt depth 1.5pt
        \hskip 1pt}}
\newcommand\bbox{\hfill \quad \blackslug \medbreak}

\newtheorem{theorem}{}[section]

\newcommand{\Proof}{\noindent{\bf Proof.}\ \ }

\tikzstyle{every node}=[circle, draw, fill=black!50, inner sep=0pt, minimum width=20pt]

\tikzstyle{input}=[circle,
                                    thick,
                                    minimum size=2.2cm,
                                    draw=black!80,
                                    fill=white!20]

\tikzstyle{input2}=[circle,
                                    thick,
                                    minimum size=1.0cm,
                                    draw=black!80,
                                    fill=white!20]

\tikzstyle{matrx}=[rectangle,
                                    thick,
                                    minimum size=1.8cm,
                                    draw=black!80,
                                    fill=white!20]

\tikzstyle{matrx2}=[rectangle,
                                    thick,
                                    minimum size=1.5cm,
                                    draw=black!80,
                                    fill=gray!20]

\tikzstyle{vecArrow} = [thick, decoration={markings,mark=at position
   1 with {\arrow[semithick]{open triangle 60}}},
   double distance=1.8pt, shorten >= 5.5pt,
   preaction = {decorate},
   postaction = {draw,line width=1.4pt, white,shorten >= 4.5pt}]
\tikzstyle{innerWhite} = [semithick, white,line width=1.4pt, shorten >= 4.5pt]

\tikzstyle{background}=[rectangle,
                                                fill=gray!10,
                                                inner sep=0.2cm,
                                                rounded corners=5mm]

\title{Coloring tournaments with forbidden substructures}

\author{
Krzysztof Choromanski \\
Google Research\\
New York, NY, USA
\and
Tony Jebara \\
Columbia University\\
New York, NY, USA
}

\date{July 28, 2012; revised \today}

\newtheorem{algorithm}{Algorithm}[section]
\newtheorem{conjecture1}{Conjecture}[section]
\newtheorem{subroutine}{Subroutine}[section]

\newcommand{\Keywords}{coloring tournaments, the Erd\H{o}s-Hajnal conjecture, transitive subtournaments}

\begin{document}
\maketitle
\begin{abstract}
Coloring graphs is an important algorithmic problem in combinatorics with many applications in computer science. In this paper we study coloring tournaments. 
A chromatic number of a random tournament is of order $\Omega(\frac{n}{\log(n)})$. The question arises whether the chromatic number can be proven to be smaller for more
structured nontrivial classes of tournaments. We analyze the class of tournaments defined by a forbidden subtournament $H$. 
This paper gives a first quasi-polynomial algorithm running in time $e^{O(\log(n)^{2})}$ that constructs colorings of $H$-free tournaments using only $O(n^{1-\epsilon(H)}\log(n))$ colors, where $\epsilon(H) \geq 2^{-2^{50|H|^{2}+1}}$ for many forbidden tournaments $H$. To the best of our knowledge all previously known related results required at least sub-exponential time and relied on the regularity lemma. Since we do not use the regularity lemma, we obtain the first known lower bounds on $\epsilon(H)$ that can be given by a closed-form expression. As a corollary, we
give a constructive proof of the celebrated open Erd\H{o}s-Hajnal conjecture with explicitly given lower bounds on the EH coefficients for all classes of prime tournaments for which the conjecture is known. Such a constractive proof was not known before. Thus we significantly reduce the gap between best lower and upper bounds on the EH coefficients from the conjecture for all known prime tournaments that satisfy it.
We also briefly explain how our methods may be used for coloring $H$-free tournaments under the following conditions: $H$ is any tournament with $\leq 5$ vertices or: $H$ is any but one tournament of six vertices. \\ 
\end{abstract}

\maketitle {\bf Keywords:} \Keywords

\section{Introduction}

Let $|\cdot|$ to denote the size of the set. Let $G$ be a graph. We denote by $V(G)$ the set of its vertices and by $E(G)$ the set of its edges. Sometimes instead of writing $|V(G)|$ we use the shorter notation $|G|$. We call $|G|$ the \textit{size of G}. For a subset $S \subseteq V(G)$ we denote by $G|S$ the subgraph of $G$ induced by $S$. A \textit{clique} in an undirected graph is a set of pairwise adjacent vertices. An \textit{independent set} in the undirected graph is a set of pairwise nonadjacent vertices. All logarithms used in this paper are of base $2$.

A \textit{tournament} is a directed graph such that, for every pair $v$ and $w$ of vertices, exactly one of the edges $(v,w)$ or $(w,v)$ exists. If $(v,w)$ is an edge in the tournament then we say that $v$ is $\textit{adjacent to}$ $w$ and $w$ is \textit{adjacent from} $v$. 
The \textit{indegree} of a vertex $v$ of a tournament $T$ is the number of vertices $w \in V(T)$ such that $(w,v) \in E(T)$.
Similarly, the \textit{outdegree} of a vertex $v$ of a tournament $T$ is the number of vertices $w \in V(T)$ such that $(v,w) \in E(T)$. 
A \textit{directed cycle} is a set of vertices $\{v_{0},...,v_{k-1}\}$ for some $k \geq 3$ such that $(v_{i},v_{(i+1) \mod k})$ is a directed edge for $i=0,...,k-1$.
A tournament is \textit{transitive} if it contains no directed cycle. For the set of vertices $V=\{v_{1},v_{2},...,v_{k}\}$ we say that an ordering $(v_{1},v_{2},...,v_{k})$ is \textit{transitive} if $v_{1}$ is adjacent to all other vertices of $V$, $v_{2}$ is adjacent to all other vertices of $V$ but $v_{1}$, etc. A subset $S \subseteq V(T)$ is \textit{transitive} if it induces a transitive tournament. 
For a tournament $H$ we say that a tournament $T$ is $H$-free if $T$ does not contain $H$ as an induced subtournament.

A \textit{proper coloring} of a tournament $T$ is an assignment of colors to its vertices such that there is no directed monochromatic cycle. Equivalently, we can consider a proper coloring of a hypergraph, where the set of vertices is $V(T)$ and the set of hyperedges consists of all triples of vertices inducing directed triangles in $T$. When properly coloring a hypergraph, we do not want to create monochromatic hyperdges. The latter is equivalent to our previous definition of coloring since one can easily note that if a monochromatic directed cycle exists in $T$ then a monochromatic directed triangle exists as well. The \textit{chromatic number} of a tournament $T$ is the minimal number of colors needed to properly color $T$.
Note that under every proper coloring of the vertices of $T$ each color class induces a transitive subtournament. Let $T^{r}_{n}$ be a random $n$-vertex tournament, where independently for every pair of vertices $\{u,v\}$, we have: $(u,v) \in E(T)$ with probability $\frac{1}{2}$. It is not hard to prove that with probability tending to $1$ as $n \rightarrow \infty$ the largest transitive subtournaments of $T^{r}_{n}$ are of logarithmic size. Thus, according to the remark above, the chromatic number of a random tournament is $\Omega(\frac{n}{\log(n)})$.

A celebrated unresolved conjecture~ of Erd\H{o}s and Hajnal states that:  

\begin{conjecture1}
\label{EHCon}
For every tournament $H$ there exists $\epsilon(H)>0$ such that every $n$-vertex $H$-free tournament contains a transitive subtournament of size at least $n^{\epsilon(H)}$.
\end{conjecture1}

In fact the conjecture was first proposed in the undirected setting by Erd\H{o}s and Hajnal but was proven to be equivalent to the directed setting above by Alon, Pach and Solymosi in 2001 (see: \cite{alon}). 
The undirected version of the conjecture (see: \cite{erdos0}) states that:

\begin{conjecture1}
\label{EHCon-undirected}
For every undirected graph $H$ there exists $\epsilon(H)>0$ such that every $n$-vertex graph $G$ that does not contain $H$ as an induced subgraph contains a clique or an independent set of size at least $n^{\epsilon(H)}$.
\end{conjecture1}

If for a given tournament $H$ there exists $\epsilon(H)>0$ then we say that \textit{$H$ satisfies the Erd\H{o}s-Hajnal conjecture with $\epsilon(H)$} or simply: \textit{$H$ satisfies the Erd\H{o}s-Hajnal conjecture}.
The coefficient $\epsilon(H)$ in the statement is called the \textit{EH coefficient}.
From now on instead of saying: "the conjecture~ of Erd\H{o}s and Hajnal" we will simply say: "the conjecture". From the context it will be always clear whether we have in mind
a directed or an undirected version.
\\
 
A subset of vertices $S \subseteq V(H)$ of a tournament $H$ is called \textit{homogeneous} if for every $v \in V(H) \backslash S$ the following holds: either $\forall_{w \in S} (w,v) \in E(H)$ or $\forall_{w \in S} (v,w) \in E(H)$. A homogeneous set $S$ is called \textit{nontrivial} if $|S|>1$ and $S \neq V(H)$. A tournament is called \textit{prime} if it does not have nontrivial homogeneous sets.

We call by \textit{$C_{5}$} a unique tournament on five vertices, where each vertex has indegree and outdegree two.

\section{Main results and related work}

Now we summarize our main results. 
We present a family of tournaments called \textit{constellations} and show a quasi-polynomial algorithm running in time $e^{(\log(n))^{2}}$ that constructs a proper coloring of an $H$-free tournament $T$, where $H$ is a constellation, with
$O(n^{1-\epsilon(H)}\log(n))$ colors, where $\epsilon(H)=2^{-2^{50|H|^{2}+1}}$ . 
We prove that:

\begin{theorem}
\label{constelcolortheorem}
If $H$ is a constellation then every $n$-vertex $H$-free tournament may be properly colored with $n^{1-\epsilon(H)}\log(n)$ colors, 
where $\epsilon(H)=\frac{1}{2^{2^{50h^{2}+1}}}$ and $h=|H|$. Furthermore, every $H$-free tournament $T$ contains a transitive
subtournament of order at least $|T|^{\epsilon(H)}$.
\end{theorem}

Constellations play important role in the conjecture since all known prime tournaments with more than six vertices satisfying the conjecture are constellations. Furthermore all tournaments for which the conjecture has been proven so far can be obtained from an infinite family of constellations and three other tournaments after applying the so-called substitution procedure introduced in \cite{alon}.
Prime tournaments are important since if the conjecture is true for prime tournaments then it is true in general.
All tournament apart from a tournament $C_{5}$ and some two six-vertex tournaments, considered in papers such as: \cite{bcc}, \cite{ccs}, \cite{seymour}, are special examples of tournaments that can be obtained from constellations using the substitution procedure.
We will briefly explain how, as a byproduct of our techniques, one can color  $H$-free tournaments with $O(n^{1-\epsilon})$ colors for every tournament $H$ on at most $5$ vertices and all but one tournament $H$ on $6$ vertices.
Our techniques give a constructive proof of the Erd\H{o}s-Hajnal conjecture for all the constellations.
Besides, after combining our methods with the substitution procedure, we obtain explicit lower bounds on EH coefficients for all known tournaments satisfying the conjecture.
The family of constellations was introduced in \cite{choromanski2}. In the same paper the conjecture was proven for them. However no algorithm to construct the $O(n^{1-\epsilon})$-coloring efficiently was given. Furthermore, even though that paper showed that an optimal coloring uses $O(n^{1-\epsilon})$ colors, the constant $\epsilon>0$ was extremely small since the proof heavily relied on the regularity lemma. 
In this paper we show that the regularity lemma is not needed at all and thus we obtain much better bounds.  
Our main contribution is an algorithmic proof that does not use the regularity lemma and a general method that can be used to show first explicit lower bounds on EH coefficients for all known tournaments satisfying the conjecture. This leads to better understanding of the asymptotics of EH coefficients. All previously known positive results regarding the conjecture for prime tournaments needed the regularity lemma. That implied big gaps between best known upper bounds on $\epsilon(H)$ of the order $\frac{1}{|H|}$ and best known lower bounds that were inversely proportional to the Szemer\'edi tower function. In this paper we significantly reduce this gap. 
In their original paper Erd\H{o}s and Hajnal asked how the EH coefficient depends on graph parameters such as its size. Our paper provides a step towards an answer for prime graphs. Previous results concerning upper bounds for EH coefficients of prime graphs were given (see: \cite{kchoromanski}, \cite{choromanski2}) but these lacked explicit lower bounds. This paper attempts to fill the gap.

Our techniques may be easily adapted to other problems regarding coloring classes of tournaments defined by forbidden subtournaments.

Let us end this section by briefly summarizing recent progress regarding the directed version of the conjecture.
\cite{seymour} described all tournaments satisfying the conjecture in the strongest, linear sense. Similarly, all tournaments satisfying the conjecture in an almost linear sense (so-called \textit{pseudocelebrities}) were described by \cite{ccs}. However, both results are for tournaments that are not prime. 
In \cite{bcc} and \cite{choromanski2} several results regarding the conjecture for prime tournaments were proven.
All the previous positive results were of purely theoretical flavor and did not easily translate into algorithmic results.

The paper is organized as follows:

\begin{itemize}
\item In Section 3 we formally define the family of constellations
      and introduce other important definitions.
\item In Section 4 we give an algorithm to color $H$-free  
      tournaments for all constellations $H$.
\item In Section 5 we
        prove correctness and analyze the running time of the
        presented algorithm and therefore prove Theorem~\ref{constelcolortheorem}.
\item In Section 6 we discuss some further applications of the introduced techniques.
\end{itemize}

\section{Constellations}

In this section we define the family of constellations and introduce other important definitions that will be used in further analysis.\\

Fix some ordering of vertices of a tournament $H$. An edge $(v,w)$ under this ordering is called a \textit{backward edge} if $w$ precedes $v$ in this ordering.
Let T be a tournament with vertex set $V(T)$ and fix some ordering of its vertices.
The \textit{graph of backward edges}  under this ordering,
denoted by $B(T, \theta)$, has vertex set $V(T)$,  and  
$\{v_i, v_j\} \in E(B(T, \theta))$ if and only if  $(v_i,v_j)$ or 
$(v_j,v_i)$ is a backward edge of $T$ under the ordering $\theta$. 
For an integer $t$, we call the graph $K_{1,t}$ a {\em star}. Let $S$ be a
star with vertex set $\{c, l_1, \ldots, l_t\}$, where $c$ is adjacent to vertices $l_1, \ldots, l_t$. We call $c$ the {\em center of the star}, and
$l_1, \ldots, l_t$ {\em the leaves of the star}. 
Note that in the case $t=1$ we may choose arbitrarily any one of the two vertices to be the center of the star, and the other vertex is then considered to be the leaf. 
Let $\theta=(v_{1},v_{2},...,v_{n})$ be an ordering of the vertex set $V(T)$ of a $n$-vertex tournament $T$. For a subset $S \subseteq V(T)$ we say that $v_{i} \in S$ is a \textit{left point of $S$ under $\theta$} if $i=\min \{j: v_{j} \in S\}$. We say that $v_{i} \in S$ is a \textit{right point of $S$ under $\theta$} if $i=\max \{j: v_{j} \in S\}$. If from the context it is clear which ordering is taken we simply say: \textit{left point of $S$} or \textit{right point of S}.  
For an ordering $\theta$ and two vertices $v_{i},v_{j}$ with $i \neq j$ we say that $v_{i}$ is \textit{before} $v_{j}$ if $i<j$ and \textit{after} $v_{j}$ otherwise.
We say that a vertex $v_{j}$ is \textit{between} two vertices $v_{i},v_{k}$ under an ordering $\theta=(v_{1},...,v_{n})$ if $i<j<k$ or $k<j<i$.

A {\em right star}
in $B(T, \theta)$ is an induced subgraph with vertex set 
$\{v_{i_0}, \ldots, v_{i_t}\}$, such that \\
$B(T,\theta)|\{v_{i_0}, \ldots, v_{i_t}\}$ is a star with center $v_{i_t}$, 
and  $i_t > i_0, \ldots, i_{t-1}$.  In this case we also  
say that $\{v_{i_0}, \ldots, v_{i_t}\}$ is  a right star in $T$.
A {\em left star}
in $B(T, \theta)$ is an induced subgraph with vertex set 
$\{v_{i_0}, \ldots, v_{i_t}\}$, such that 
$B(T,\theta)|\{v_{i_0}, \ldots, v_{i_t}\}$ is a star with center $v_{i_0}$, 
and  $i_0 < i_1, \ldots, i_t$.    In this case we also  
say that $\{v_{i_0}, \ldots, v_{i_t}\}$ is a  left star in $T$.
From now on whenever we will refer to the star in $B(T, \theta)$ we will mean a left star or a right star.

Let $H$ be a tournament and assume there is an ordering $\theta$ of its vertices such that every connected component of $B(H, \theta)$ is  either a star or a singleton under this ordering. We call this ordering a \textit{star ordering}. The \textit{interstellar graph} of $H$ under a star ordering $\theta$ is an undirected graph, whose vertices are the sets of leaves of the stars of $H$ under $\theta$ and any two given vertices $L_{1}$ and $L_{2}$ are adjacent iff:
\begin{itemize}
\item the left point of $L_{1}$ precedes the right point of $L_{2}$ in $\theta$ and
\item the left point of $L_{2}$ precedes the right point of $L_{1}$ in $\theta$.
\end{itemize}    

\begin{center}
\begin{tikzpicture}[every path/.style={>=latex},every node/.style={draw,circle}]
\def \n {5}
\def \radius {3cm}

  \node            (a) at (1.5,0)  { 1 };
  \node            (b) at (3,0)  { 2 };
  \node            (c) at (4.5,0)  { 3 };
  \node            (d) at (6,0)  { 4 };
  \node            (e) at (7.5,0)  { 5 };
  \node            (f) at (9,0)  { 6 };
  \node            (g) at (10.5,0)  { 7 };
  \node            (h) at (12,0)  { 8 };
  \node            (i) at (13.5,0)  { 9 };
  \node            (j) at (15,0)  { 10 };
  \node            (k) at (16.5,0)  { 11 };

 \draw[->] (f) edge [out=145,in=35] (a);
 \draw[->] (c) edge [out=150,in=30] (a);
 \draw[->] (j) edge [out=150,in=30] (b);
 \draw[->] (j) edge [out=155,in=25] (e);
 \draw[->] (k) edge [out=225,in=315] (d);
 \draw[->] (k) edge [out=220,in=320] (g);
 \draw[->] (k) edge [out=215,in=325] (i);

\end{tikzpicture}
\end{center}

\begin{center}
\begin{tikzpicture}[every path/.style={>=latex},every node/.style={draw,circle}]
\def \n {3}
\def \radius {3cm}
\def \margin {10}

  \node[draw, circle] (a) at ({360/\n * (1 - 1)+90}:\radius) {$\Sigma_{1}$};
  \node[draw, circle] (b) at ({360/\n * (2 - 1)+90}:\radius) {$\Sigma_{2}$};
  \node[draw, circle] (c) at ({360/\n * (3 - 1)+90}:\radius) {$\Sigma_{3}$};

  \draw (a) edge  (b);
  \draw (a) edge  (c);
  \draw (b) edge  (c);

\end{tikzpicture}
\end{center}
\begin{center} 
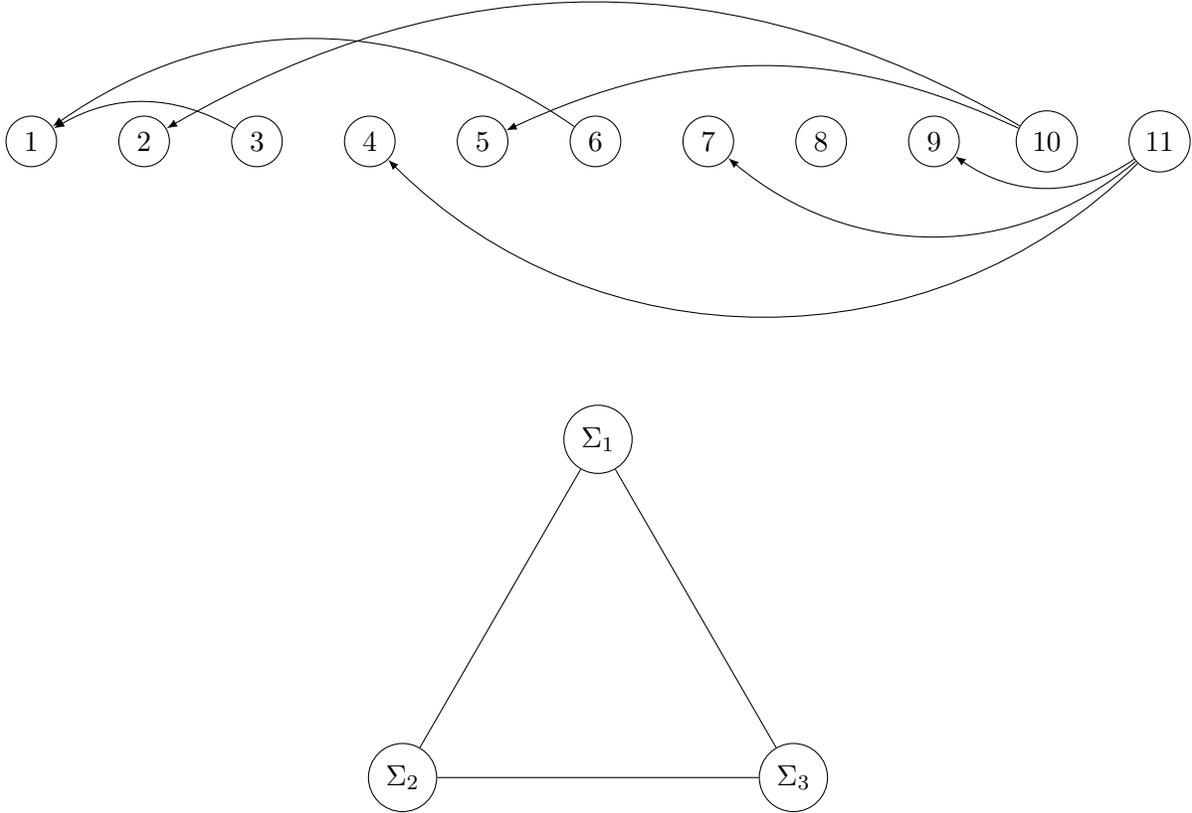
Fig.1  Constellation with backward edges drawn under its constellation ordering and consisting of three stars: $\Sigma_{1},\Sigma_{2},\Sigma_{3}$  (above).
The interstellar graph where nodes correspond to the stars and two edges are adjacent if and only if the intervals defined by the sets of leaves under given constellation
ordering intersect (below). 
\end{center}

For each connected component $C$ of the interstellar graph of $H$ denote by $Z(C)$ the union of subsets of $V(H)$ corresponding to its vertices (this is the union of some subsets of $V(H)$ of the vertices of $H$). 
Next let us define $\mathcal{C}({Z(C)})$ as follows.
We say that a vertex $v \in \mathcal{C}({Z(C)})$ if $v \in Z(C)$ or $v$ is between some two vertices of $Z(C)$ under the ordering $\theta$.
Let $C_{1},...,C_{k}$ be the connected components of the interstellar graph. Note that for any given $1 \leq i < j \leq k$ either every vertex of $\mathcal{C}({Z(C_{i})})$ is before every vertex of $\mathcal{C}({Z(C_{j})})$, or every vertex of $\mathcal{C}({Z(C_{j})})$ is before every vertex of $\mathcal{C}({Z(C_{i})})$. Thus there is a natural ordering of the sets $\mathcal{C}({Z(C_{i})})$ for $i=1,2,...,k$ induced by the ordering of the vertices. Denote the ordered sequence of the sets $\mathcal{C}({Z(C_{i})})$ for $i=1,2,...,k$  as 
$(\mathcal{W}_{1},...,,\mathcal{W}_{k})$, where a set $\mathcal{W}_{i}$ is before a set $\mathcal{W}_{j}$ for $1 \leq i < j \leq k$. 
Denote $\mathcal{W}_{0}=\mathcal{W}_{k+1} = \emptyset$.
For $i=1,2,...,k+1$ denote by $\mathcal{R}_{i}$  the set of the vertices of $H$ that are after all the vertices of $\mathcal{W}_{i-1}$ and before all the vertices of $\mathcal{W}_{i}$ under the ordering $\theta$.  
Note that if $\mathcal{R}_{i}$ is nonempty then all its elements are centers of the stars of $H$.
Denote the set of nonempty sets $\mathcal{R}_{i}$ as $\{\mathcal{M}_{1},...,\mathcal{M}_{r}\}$ for some $r \geq 0$.
Note that $\{\mathcal{W}_{1},...,\mathcal{W}_{k},\mathcal{M}_{1},...,\mathcal{M}_{r}\}$  is a partition of the vertices of $H$.
Denote this partition by $P_{\theta}(H)$.
We are ready to define constellations.
 
A tournament $T$ is a \textit{constellation} if there exists a star ordering $\theta$ of its vertices such that if a center of a star belongs to some set $P \in P_{\theta}(H)$ then no leaf of this star belongs to $P$.\\

We call such an ordering a {\em constellation ordering} of $T$. Let 
$\Sigma_1, \ldots, \Sigma_l$ be the non-singleton components of 
$B(T, \theta)$. We say that $\Sigma_1, \ldots, \Sigma_l$ 
are the {\em stars of $T$ under} $\theta$. If $V(T)=\bigcup_{i=1}^l V(\Sigma_l)$, 
we say that $T$ is a {\em regular constellation}.

Even though the definition of the family of constellations that we have just presented seems to be complicated, it is in fact easy to construct examples of constellations of an arbitrary size. That is because our definition uses the notion of the constellation ordering and this term has natural and straightforward pictorial interpretation.
\begin{center}
\begin{tikzpicture}[every path/.style={>=latex},every node/.style={draw,circle}]
\def \n {5}
\def \radius {3cm}

  \node            (a) at (1,0)  { 0 };
  \node            (b) at (2.5,0)  { 1 };
  \node            (c) at (4,0)  { 2 };
  \node            (d) at (5.5,0)  { 3 };
  \node            (e) at (7,0)  { 4 };
  \node            (f) at (8.5,0)  { 5 };
  \node            (g) at (10,0)  { 6 };
  \node            (h) at (11.5,0)  { 7 };
  \node            (i) at (13,0)  { 8 };
  \node            (j) at (14.5,0)  { 9 };

  \draw[->] (f) edge [out=140,in=40] (a);
  \draw[->] (f) edge [out=145,in=35] (c);
  \draw[->] (f) edge [out=160,in=20] (e);
  \draw[->] (g) edge [out=210,in=330] (d);
  \draw[->] (i) edge [out=220,in=320] (d);
  \draw[->] (j) edge [out=230,in=310] (d);

\end{tikzpicture}
\end{center}

\begin{center} 
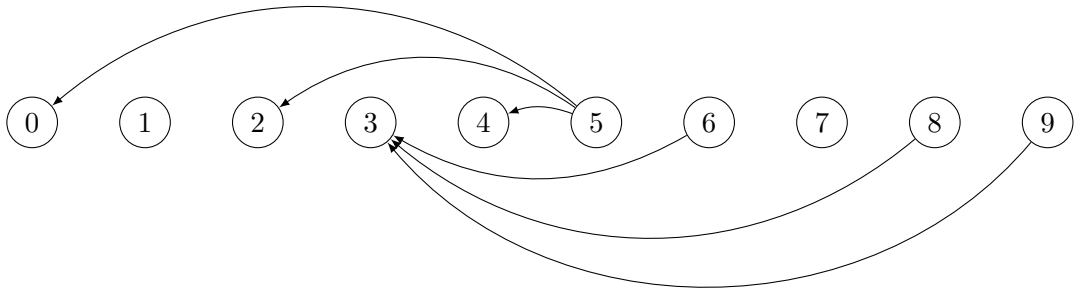
Fig.2 Constellation consisting of two stars - one left and one right. For the clarity of the picture only backward edges were drawn.
\end{center}

A \textit{galaxy ordering} of the vertices of the tournament is the constellation ordering under which no center of the star appears between leaves of another star.
Notice that this is not necessarily the case for the constellation ordering. A center of the star can be between leaves (call this set $\mathcal{L}$) of another star, but if it happens then all its leaves have to be
in a different connected component of the interstellar graph that the one that corresponds to $\mathcal{L}$.
A \textit{galaxy} is a tournament that has a galaxy ordering of vertices. Galaxies is a subfamily of constellations. The conjecture was proven for them in 
\cite{bcc}. That was the first result where the conjecture was proven for an infinite family of prime tournaments. Our constructive proof gives much better
lower bounds on EH coefficients for galaxies than those from \cite{bcc}.

\begin{center}
\begin{tikzpicture}[every path/.style={>=latex},every node/.style={draw,circle}]
\def \n {5}
\def \radius {3cm}

  \node            (a) at (2,0)  { 1 };
  \node            (b) at (4,0)  { 2 };
  \node            (c) at (6,0)  { 3 };
  \node            (d) at (8,0)  { 4 };
  \node            (e) at (10,0)  { 5 };
  \node            (f) at (12,0)  { 6 };
  \node            (g) at (14,0)  { 7 };
  \node            (h) at (16,0)  { 8 };

  \draw[->] (c) edge [out=160,in=20] (a);
  \draw[->] (e) edge [out=150,in=30] (a);
  \draw[->] (h) edge [out=120,in=60] (f);
  \draw[->] (g) edge [out=200,in=340] (d);
  \draw[->] (g) edge [out=210,in=330] (b);

\end{tikzpicture}
\end{center}

\begin{center} 
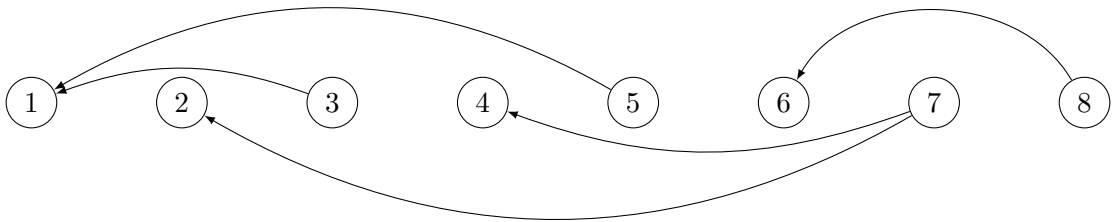
Fig.3 Galaxy consisting of one left and two right stars. All edges that are not drawn are forward.
\end{center}

\subsection{Some useful definitions}

This section provides several definitions used in the paper.

Take a tournament $T$. 
Let $X,Y \subseteq V(T)$ be disjoint, where $|X|,|Y|>0$. Denote by $e_{X,Y}$ the number of directed edges $(x,y)$, where $x \in X$ and $y \in Y$.
The \textit{directed density from X to Y} is defined as 
$d(X,Y)=\frac{e_{X,Y}}{|X||Y|}.$

We say that a tournament $T$ is \textit{$\epsilon$-transitive} if it contains a transitive subtournament of order at least $|T|^{\epsilon}$.

For the transitive pairwise disjoint subsets $T_{1},T_{2},...,T_{k} \subseteq V(T)$ we say that a sequence \\$(T_{1},T_{2},...,T_{k})$ is a \textit{$(c,\lambda, \epsilon)$-$t$-sequence  of length $k$} (where \textit{t} stands for: \textit{transitive}) if the following holds: 
\begin{itemize}
\item $d(T_{i},T_{j}) \geq 1 - \lambda$ for $1 \leq i < j \leq k$, and
\item $|T_{i}| \geq c|T|^{\epsilon}$ for $i=1,2,...,k$.
\end{itemize}

For the transitive pairwise disjoint subsets $T_{1},T_{2},...,T_{k} \subseteq V(T)$ we say that a sequence \\$(T_{1},T_{2},...,T_{k})$ is a \textit{smooth $(c,\lambda, \epsilon)$-$t$-sequence of length $k$} if the following holds: 
\begin{itemize}
\item $d(\{v\},T_{j}) \geq 1 - \lambda$ for $1 \leq i < j \leq k$ and $v\in T_{i}$,
\item $d(T_{i},\{v\}) \geq 1 - \lambda$ for $1 \leq i < j \leq k$ and $v\in T_{j}$, and
\item $|T_{i}| \geq c|T|^{\epsilon}$ for $i=1,2,...,k$.
\end{itemize}

Note that every smooth $(c,\lambda, \epsilon)$-$t$-sequence is a $(c,\lambda, \epsilon)$-$t$-sequence.

For the pairwise disjoint subsets $A_{1},...,A_{k}\subseteq V(T)$ we say that a sequence $(A_{1},A_{2},...,A_{k})$ is a \textit{$(c,\lambda)$-$l$-sequence of length $k$} (where \textit{l} stands for: \textit{linear}) if the following holds:
\begin{itemize}
\item $d(A_{i},A_{j}) \geq 1 - \lambda$ for $1 \leq i < j \leq k$, and
\item $|A_{i}| \geq c|T|$ for $i=1,2,...,k$.
\end{itemize}

For the pairwise disjoint subsets $A_{1},...,A_{k}\subseteq V(T)$ we say that a sequence $(A_{1},A_{2},...,A_{k})$ is a \textit{smooth $(c,\lambda)$-$l$-sequence of length $k$} if the following holds:
\begin{itemize}
\item $d(\{v\},A_{j}) \geq 1 - \lambda$ for $1 \leq i < j \leq k$ and $v \in A_{i}$,
\item $d(A_{i},\{v\}) \geq 1 - \lambda$ for $1 \leq i < j \leq k$ and $v \in A_{j}$, and
\item $|A_{i}| \geq c|T|$ for $i=1,2,...,k$.
\end{itemize}

Note that every smooth $(c,\lambda)$-$l$-sequence is a $(c,\lambda)$-$l$-sequence.

For the pairwise disjoint subsets $A_{1},...,A_{k+1},T_{1},...,T_{k} \subseteq V(T)$, where $T_{1},...,T_{k}$ are transitive, we say that a sequence 
$(A_{1},T_{1},...,A_{k},T_{k},A_{k+1})$ is a \textit{$(c,\lambda,\epsilon)$-$m$-sequence of length $2k+1$} (where \textit{m} stands for: \textit{mixed}) if the following holds:
\begin{itemize}
\item $d(T_{i},T_{j}) \geq 1 - \lambda$ for $1 \leq i < j \leq k$,
\item $d(A_{i},A_{j}) \geq 1 - \lambda$ for $1 \leq i < j \leq k$,
\item $d(A_{i},T_{j}) \geq 1 - \lambda$ for $1 \leq i \leq j \leq k$,
\item $d(T_{i},A_{j}) \geq 1 - \lambda$ for $1 \leq i < j \leq k$,
\item $|A_{i}| \geq c|T|$ for $i=1,2,...,k,k+1$, and
\item $|T_{i}| \geq c|T|^{\epsilon}$ for $i=1,2,...,k$.
\end{itemize}

We refer to the sets $T_{1},...,T_{k}$ from the $m$-sequence as \textit{transitive sets of the $m$-sequence}. 
We say that a $m$-sequence of length $2k+1$ is \textit{$M$-big} if $T_{i} \geq M$ for $i=1,2,...,k$.

\begin{center}
\begin{tikzpicture}[every path/.style={>=latex},every node/.style={draw,circle}]
\def \n {5}
\def \radius {3cm}

  \node[input]            (a) at (3.3,0)  { $A_{1}$ };
  \node[matrx]            (b) at (6.6,0)  { $T_{1}$ };
  \node[input]            (c) at (9.9,0)  { $A_{2}$ };
  \node[matrx]            (d) at (13.5,0)  { $T_{2}$ };
  \node[input]            (e) at (16.8,0)  { $A_{3}$ };

  \draw[vecArrow] (a) to (b);
  \draw[innerWhite] (a) to (b);

  \draw[vecArrow] (b) to (c);
  \draw[innerWhite] (b) to (c);

  \draw[vecArrow] (c) to (d);
  \draw[innerWhite] (c) to (d);

  \draw[vecArrow] (d) to (e);
  \draw[innerWhite] (d) to (e);

\end{tikzpicture}
\end{center}

\begin{center} 
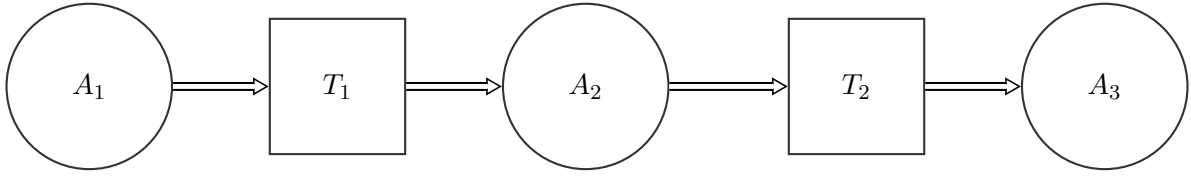
Fig.4 Schematical representation of the $(c,\lambda,\epsilon)$-$m$-sequence. This sequence consists of three linear sets: $A_{1},A_{2},A_{3}$ and two transitive sets: $T_{1}$ and $T_{2}$. The arrows indicate the orientation of most
of the edges going between different elements of the $(c,\lambda, \epsilon)$-$m$-sequence. Each $T_{i}$ satisfies: $|T_{i}| \geq c n^{\epsilon}$ and each $A_{i}$ satisfies: $|A_{i}| \geq c \cdot n$, where $n=|T|$. 
\end{center}

For the pairwise disjoint subsets $A_{1},...,A_{k+1},T_{1},...,T_{k} \subseteq V(T)$, where $T_{1},...,T_{k}$ are transitive, we say that a sequence 
$(A_{1},T_{1},...,A_{k},T_{k},A_{k+1})$ is a \textit{smooth $(c,\lambda,\epsilon)$-$m$-sequence of length $2k+1$} if the following holds:
\begin{itemize}
\item sequence $(T_{1},...,T_{k})$ is a smooth $(c,\lambda,\epsilon)$-$t$-sequence
\item sequence $(A_{1},...,A_{k+1})$ is a smooth $(c,\lambda)$-$l$-sequence 
\item $d(\{v\},T_{j}) \geq 1 - \lambda$ for $1 \leq i \leq j \leq k$ and $v \in A_{i}$,
\item $d(A_{i},\{v\}) \geq 1 - \lambda$ for $1 \leq i \leq j \leq k$ and $v \in T_{j}$,
\item $d(\{v\},A_{j}) \geq 1 - \lambda$ for $1 \leq i < j \leq k$ and $v \in T_{i}$,
\item $d(T_{i},\{v\}) \geq 1 - \lambda$ for $1 \leq i < j \leq k$ and $v \in A_{j}$,
\item $|A_{i}| \geq c|T|$ for $i=1,2,...,k,k+1$, and
\item $|T_{i}| \geq c|T|^{\epsilon}$ for $i=1,2,...,k$.
\end{itemize}

Note that every smooth $(c,\lambda,\epsilon)$-$m$-sequence is a $(c,\lambda,\epsilon)$-$m$-sequence.

If a smooth $(c,\lambda,\epsilon)$-$m$-sequence $(A_{1},T_{1},...,A_{k},T_{k},A_{k+1})$ satisfies:

\begin{itemize}
\item $d(T_{i},T_{j}) = 1$ for $i,j \in \mathcal{I}$, where $i<j$ and $\mathcal{I} \subseteq \{1,2,...,k\}$,
\end{itemize}

then we say that $(A_{1},T_{1},...,A_{k},T_{k},A_{k+1})$ is an \textit{$\mathcal{I}$-strong $(c,\lambda,\epsilon)$-$m$-sequence of length $2k+1$}.  

Whenever we do not care about parameters of the $(c,\lambda,\epsilon)$-$t$-sequences, $(c,\lambda)$-$l$-sequences or $(c,\lambda,\epsilon)$-$m$-sequences under consideration, we refer to them simply as: $t$-sequences, $l$-sequences and $m$-sequences respectively.

For two disjoint subsets $A_{1},T_{1} \subseteq V(T)$ such that $T_{1}$ is transitive we say that a pair $(A_{1},T_{1})$ is \textit{$(c,\epsilon)$-saturated} if the following holds:
\begin{itemize}
\item $|A_{1}| \geq c|T|$,
\item $|T_{1}| \geq c|T|^{\epsilon}$, and
\item $d(A_{1},T_{1})=1$ or $d(T_{1},A_{1})=1$.
\end{itemize}

\section{Quasi-polynomial algorithm for coloring $H$-free tournaments for a constellation $H$}

\subsection{Introduction}

Whenever we consider algorithms involving tournaments, we assume that the input is a tournament description given by adjacency lists.
This section presents two main algorithms (and several subroutines used by them): one that finds a polynomial-size transitive subtournament in the strong $m$-sequence of an $H$-free tournament (algorithm \textit{PolyTrans}) and one that colors an $H$-free tournament (algorithm \textit{Color-H-free}), where $H$ is a constellation. The former is in fact used to construct the coloring produced by the latter one. 
The latter one takes as an input only forbidden constellation $H$ and $H$-free tournament $T$.
The core part of the algorithm \textit{PolyTrans} is the recursive algorithm \textit{PolyTransCore} that gets as an input colored $m$-sequence (the coloring is done in the initial phase of the algorithm \textit{PolyTrans}) and outputs a polynomial-size transitive subset.

We would like to give a general idea of our algorithmic approach first since the algorithm is complicated. Assuming that $T$ is an $H$-free tournament, we find in $T$ a long enough sequence of big linear and transitive sets such that most of the edges between those sets go from these sets that are earlier in the sequence to those that are placed later. At that point no assumption about the structure of $H$ is required. We handle that part of the algorithm without the use of the regularity lemma and that enables us to significantly improve best known lower bounds on $\epsilon(H)$. In the second stage we heavily use the fact that $H$, as a constellation, has a specific ordering of vertices. 
The algorithm finding polynomial-size transitive subtournament uses the following technical procedures: \textit{Find-L-Sequence}, \textit{Find-Clique}, \textit{Find-M-Sequence}, \textit{MakeSmooth}, \textit{FindStrong-M-Sequence}.
Algorithm \textit{Find-L-Sequence} finds a $(c,\lambda)$-$l$-sequence in an $H$-free tournament.
Algorithm \textit{Find-Clique} finds a clique in a dense $k$-partite undirected graph.
Algorithm \textit{Find-M-Sequence} is responsible for finding $m$-sequences in $H$-free tournaments. Algorithm \textit{MakeSmooth} makes them smooth. Finally, algorithm \textit{FindStrong-M-Sequence} constructs a strong $m$-sequence in an $H$-free tournament. That sequence is the input for the \textit{PolyTrans} algorithm. This short summary will be clearer later when we describe all the algorithms in detail.

Let $H$ be a constellation with $h=|H|$. Let $\theta$ be a constellation ordering of $H$.
We define a function $\zeta : V(H) \rightarrow \mathbb{N}$ as follows:
\begin{itemize}
\item if $v \in V(H)$ is a leaf of a star let $\zeta(v)=i(2h+1)+1$, where $v$ is the $i$th leaf under ordering $\theta$,
\item otherwise let $\zeta(v)=j(2h+1) + 2r$, where $v$ is the $r$th center after the $j$th leaf and before the $(j+1)$th leaf under the ordering $\theta$.
\end{itemize}

First we show algorithms: \textit{PolyTrans} and \textit{Color-H-free}. Then we will describe all supportive procedures mentioned above.

\subsection{An overview of the method}

The constructive proof we are about to present that all the constellations satisfy the conjecture is very technical. 
Therefore before going into details we would
like to explain main steps of the proof. In this subsection we give reader the intuition how the proof works, what are the most important
parts of the proof, finally - why the ideas used to prove the conjecture for galaxies are not sufficient to succeed
with constellations and what are the new techniques that need to be used in this setting.

Notice first that we can always assume that we have a sequence of linear sets and big transitive subtournaments such that
for any two of them the one that appears first in the sequence is almost complete to the one that appears later.
This sequence is what we call a $(c,\lambda, \epsilon)$-$m$-sequence. Parameter $c$ encodes lower bounds on the sizes of linear and
transitive elements. Parameter $\epsilon$ specifies the value of the exponent in the lower bound on the size of the transitive element.
Finally, parameter $\lambda$ specifies lower bound on the directed density between different elements of the sequence. 
By "almost adjacent" we mean that the directed density is very close to $1$. 
Existence of such a sequence is an immediate consequence of the regularity lemma, but since we do not want to use that tool, we prove that fact
in a very different way. This enables us to get a bound on the EH coefficient that can be expressed in the compact way.
What is important here is that at this point we did not need
to assume any specific structure of the tournament $H$ that is being excluded. 
If we need to summarize the proof of the conjecture for galaxies from \cite{bcc} in one sentence, we should say:
"linear sets for centers of stars, transitive subtournaments for leaves". We give a proof by contradiction,
assume that an $H$-free tournament does not have polynomial-size transitive subtournaments and construct a copy of $H$
in it. The copy will be constructed star by star. We proceed by creating first an appropriate 
$(c,\lambda,\epsilon)$-$m$-sequence mentioned above. 
Then we are looking for stars with centers in linear sets and leaves in transitive sets. 
When the star is found it is removed and the entire sequence is updated. The update is
done in such a wat that we can simply merge a star we have just found with the remaining part of the tournament $H$ (that will
be found in the new sequence) to get a copy of $H$.

\begin{center}
\begin{tikzpicture}[every path/.style={>=latex},every node/.style={draw,rectangle}]
\def \n {5}
\def \radius {3cm}
\def \sh {1cm}

  \node (a) [input2] at (2.5,3.5) {$\mathbf{w}_{1}$};
  \node (b) [input2] at (2.5,2.0) {$\mathbf{w}_{2}$};
  \node (c) [input2] at (2.5,0.5) {$\mathbf{w}_{3}$};

  \node (d) [input2] at (7.5,3.5) {$\mathbf{u}_{1}$};
  \node (e) [input2] at (7.5,2.0) {$\mathbf{u}_{2}$};
  \node (f) [input2] at (7.5,0.5) {$\mathbf{u}_{3}$};

  \node (g) [input2] at (12.5,5.0) {$\mathbf{w}_{1}$};
  \node (h) [input2] at (12.5,3.75) {$\mathbf{w}_{2}$};
  \node (i) [input2] at (12.5,2.5) {$\mathbf{w}_{3}$};
  \node (j) [input2] at (12.5,1.25) {$\mathbf{u}_{1}$};
  \node (k) [input2] at (12.5,.0) {$\mathbf{u}_{2}$};
  \node (l) [input2] at (12.5,-1.25) {$\mathbf{u}_{3}$};

\draw (2,2) circle (2.5cm);

  \draw[->] (a) edge  (d);
  \draw[->] (a) edge  (e);
  \draw[->] (a) edge  (f);
  \draw[->] (b) edge  (d);
  \draw[->] (b) edge  (e);
  \draw[->] (b) edge  (f);
  \draw[->] (c) edge  (d);
  \draw[->] (c) edge  (e);
  \draw[->] (c) edge  (f);

  \draw[->] (a) edge  (b);
  \draw[->] (b) edge  (c);
  \draw[->] (d) edge  (e);
  \draw[->] (e) edge  (f);

  \draw[->] (g) edge  (h);
  \draw[->] (h) edge  (i);
  \draw[->] (i) edge  (j);
  \draw[->] (j) edge  (k);
  \draw[->] (k) edge  (l);

 \draw[->] (a) edge [out=210,in=150] (c);
 \draw[->] (d) edge [out=330,in=30] (f);

\begin{pgfonlayer}{background}
    \node [background,
                fit=(a) (c)] {};
\end{pgfonlayer}

\begin{pgfonlayer}{background}
    \node [background,
                fit=(d) (f)] {};
\end{pgfonlayer}

\begin{pgfonlayer}{background}
    \node [background,
                fit=(g) (l)] {};
\end{pgfonlayer}

\end{tikzpicture}
\end{center}

\begin{center} 
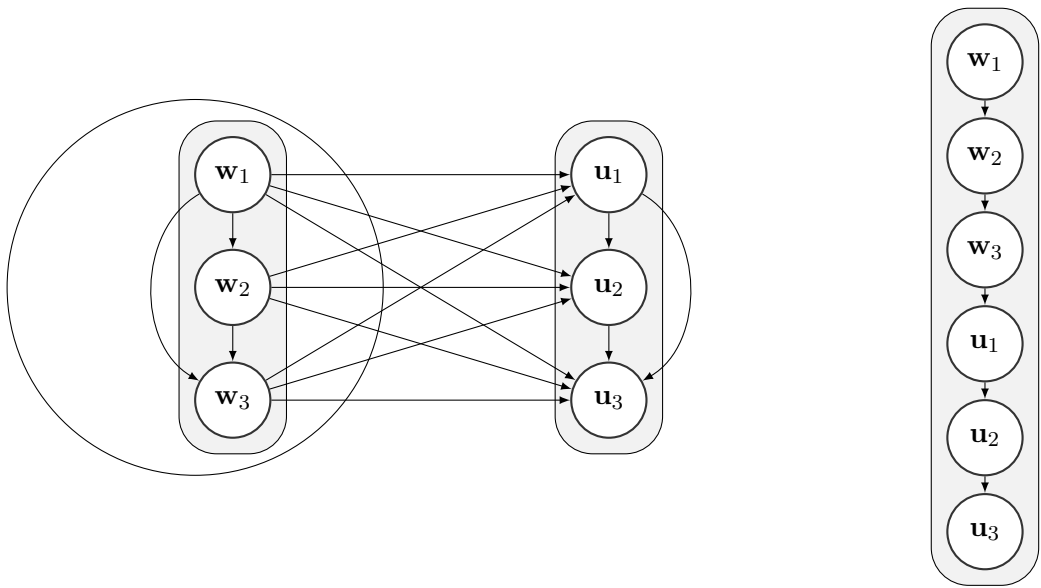
Fig.5 The method to construct polynomial-size transitive subtournament. A circle represents a linear set and 
$\{w_{1},...,w_{3}\}$ is a big transitive subset that (by induction) can be extracted from it. A set $\{u_{1},...,u_{3}\}$
represents another transitive set of substantial size. Set $\{w_{1},...,w_{3}\}$ is complete to $\{u_{1},...,u_{3}\}$.
Thus $\{w_{1},...,w_{3},u_{1},...,u_{3}\}$ is also a transitive set. For $\epsilon>0$ small enough this set is of size at least $n^{\epsilon}$, where $n$ is the size of the $H$-free tournament $T$.
\end{center}

If the star cannot be found, then a simple argument using Pigeongole principle shows that we must have a substantial transitive subset
complete to/from a linear set (see: Fig. 5). The key observation here is that a transitive set of the substantial size complete to or from a linear set gives
us a polynomial-size transitive subtournament.
The crucial element that makes this method work is that whenever we are looking for a star there is no need to take care
of the right type of adjacency between leaves if all candidates for them were chosen from the same transitive set.
The right type of adjacency is given for granted. The price that is paid is the fact that we cannot allow
centers of stars to be between leaves of another star since that would require looking for centers also in transitive chunks.
That in turn will not enable us to get a polynomial-size transitive subtournament.
This was also the main reason why the authors started to work on more general techniques that would handle cases where centers of stars are between leaves of another stars, i.e. more general configurations of stars.
The conceptual idea of this more general technique is to change the paradigm: "centers in linear sets, leaves in transitive subtournaments" by starting
with the $(c, \lambda, w)$-$t$-sequence that consists only of big transitive chunks (by big we mean of size
at least $c \cdot |T|^{\epsilon}$ for some constant $c, \epsilon>0$).

We start proceeding as in the galaxy-proof, but try to use $t$-sequences first
(in particular we look for centers of stars in transitive chunks, see Fig. 6). 


\begin{center}
\begin{tikzpicture}[every path/.style={>=latex},every node/.style={draw,rectangle}]
\def \n {5}
\def \radius {3cm}

  \node (a) [input2] at (2,0) {$\mathbf{v}_{1}$};
  \node (b) [input2] at (4,0) {$\mathbf{v}_{2}$};
  \node (c) [input2] at (6,0) {$\mathbf{v}_{3}$};
  \node (d) [input2] at (8,0) {$\mathbf{v}_{4}$};
  \node (e) [input2] at (10,0) {$\mathbf{v}_{5}$};
  \node (f) [input2] at (12,0) {$\mathbf{v}_{6}$};
  \node (g) [input2] at (14,0) {$\mathbf{v}_{7}$};

 \draw[->] (e) edge [out=210,in=330] (a);
 \draw[->] (e) edge [out=205,in=335] (c);
 \draw[->] (g) edge [out=150,in=30] (d);
 \draw[->] (g) edge [out=165,in=15] (f);

\begin{pgfonlayer}{background}
    \node [background,
                fit=(a) (c)] {};
\end{pgfonlayer}

\begin{pgfonlayer}{background}
    \node [background,
                fit=(d) (f)] {};
\end{pgfonlayer}

\begin{pgfonlayer}{background}
    \node [background,
                fit=(g) (g)] {};
\end{pgfonlayer}


\end{tikzpicture}
\end{center}


\begin{center}
\begin{tikzpicture}[every path/.style={>=latex},every node/.style={draw,rectangle}]
\def \n {5}
\def \radius {3cm}

  \node (a) [matrx2] at (2,0) {$\mathbf{Tr}_{1}$};
  \node (b) [matrx2] at (3.5,0) {$\mathbf{Tr}_{2}$};
  \node (c) [matrx2] at (5.0,0) {$\mathbf{Tr}_{3}$};

  \node (d) [matrx2] at (8.0,0) {$\mathbf{Tr}_{4}$};
  \node (e) [matrx2] at (9.5,0) {$\mathbf{Tr}_{5}$};
  \node (f) [matrx2] at (11.0,0) {$\mathbf{Tr}_{6}$};

  \node (g) [matrx2] at (14.0,0) {$\mathbf{Tr}_{7}$};

  \draw[vecArrow] (c) to (d);
  \draw[innerWhite] (c) to (d);

  \draw[vecArrow] (f) to (g);
  \draw[innerWhite] (f) to (g);

\end{tikzpicture}
\end{center}

\begin{center} 
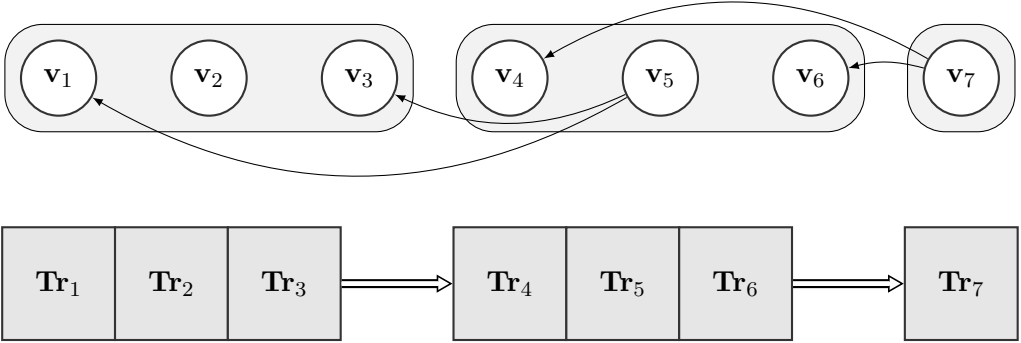
Fig.6  Constellation $H$ with its constellation ordering of vertices (above) and the associated $t$-sequence $\chi$. Sequence $\chi$ consists of three transitive sets, first two are partitioned into three equal-length subsets.
There is a 1-1 mapping between vertices of $H$ and subchunks $Tr_{i}$. The goal is to look for a node $v_{i}$ in $Tr_{i}$.
\end{center}

The problem we face (that was mentioned by us before) is that now we do not necessarily get as an outcome a big linear set
complete to or from a big transitive chunk. Instead, we obtain two transitive chunks such that one of them is
complete to the other one. However those chunks, even after merging, may not give big
enough transitive subtournament. This is the place where we need to use strong $(c,\lambda,\epsilon)$-$m$-sequences.

We repeat our previous procedure several times in many different regions of the long enough $t$-sequence. 
By doing it, using Ramsey argument, we can conclude that we get arbitrarily large set of transitive
subchunks of the elements of the sequence with the additional property that the subchunks appearing earlier in the
sequence are complete to those appearing later (altogether they may still not give a big enough transitive
subtournament). We may also assume without loss of generality that between those subchunks we have big linear sets
such that each linear set is almost complete from all subchunks preceeding it and almost complete to all subchunks
that it preceeds. This can be easily done if we slightly enrich the $t$-sequence we started with
by introducing linear sets between transitive chunks. This can be always achieved since we have already observed
that we can start with the arbitrarily long $(c,\lambda,\epsilon)$-$m$-sequence.
But now we can again try to build a constellation star by star. 
Notice however that in contrast to the previous scenario, we can look for different leaves of the same star in different
transitive chunks (see Fig. 7). This is possible because for any two transitive chunks of the strong
sequence the one appearing earlier in the sequence is complete to the one appearing later. Therefore while
constructing a star in such a way that leaves are being found in transitive chunks we have the right type
of adjacency between those candidates for leaves for granted. This is no longer true if the structure we are given
is no strong. 


\begin{center}
\begin{tikzpicture}[every path/.style={>=latex},every node/.style={draw,rectangle}]
\def \n {5}
\def \radius {3cm}

  \node (a) [input2] at (2,0) {$\mathbf{v}_{1}$};
  \node (b) [input2] at (4,0) {$\mathbf{v}_{2}$};
  \node (c) [input2] at (6,0) {$\mathbf{v}_{3}$};
  \node (d) [input2] at (8,0) {$\mathbf{v}_{4}$};
  \node (e) [input2] at (10,0) {$\mathbf{v}_{5}$};
  \node (f) [input2] at (12,0) {$\mathbf{v}_{6}$};
  \node (g) [input2] at (14,0) {$\mathbf{v}_{7}$};

 \draw[->] (e) edge [out=210,in=330] (a);
 \draw[->] (e) edge [out=205,in=335] (c);
 \draw[->] (g) edge [out=150,in=30] (d);
 \draw[->] (g) edge [out=165,in=15] (f);

\begin{pgfonlayer}{background}
    \node [background,
                fit=(a) (a)] {};
\end{pgfonlayer}

\begin{pgfonlayer}{background}
    \node [background,
                fit=(c) (d)] {};
\end{pgfonlayer}

\begin{pgfonlayer}{background}
    \node [background,
                fit=(f) (f)] {};
\end{pgfonlayer}


\end{tikzpicture}
\end{center}

\begin{center}
\begin{tikzpicture}[every path/.style={>=latex},every node/.style={draw,rectangle}]
\def \n {5}
\def \radius {3cm}

  \node (a) [matrx2] at (2,0) {$\mathbf{Tr}_{1}$};
  \node (b) [input] at (4.5,0) {$\mathbf{A}_{1}$};
  \node (c) [matrx2] at (7.0,0) {$\mathbf{Tr}_{2}$};
  \node (d) [matrx2] at (8.5,0) {$\mathbf{Tr}_{3}$};
  \node (e) [input] at (11,0) {$\mathbf{A}_{2}$};

  \node (f) [matrx2] at (13.5,0) {$\mathbf{Tr}_{4}$};
  \node (g) [input] at (16,0) {$\mathbf{A}_{3}$};

  \draw[vecArrow] (a) to (b);
  \draw[innerWhite] (a) to (b);
  \draw[vecArrow] (b) to (c);
  \draw[innerWhite] (b) to (c);
  \draw[vecArrow] (d) to (e);
  \draw[innerWhite] (d) to (e);
  \draw[vecArrow] (e) to (f);
  \draw[innerWhite] (e) to (f);
  \draw[vecArrow] (f) to (g);
  \draw[innerWhite] (f) to (g);

\end{tikzpicture}
\end{center}

\begin{center} 
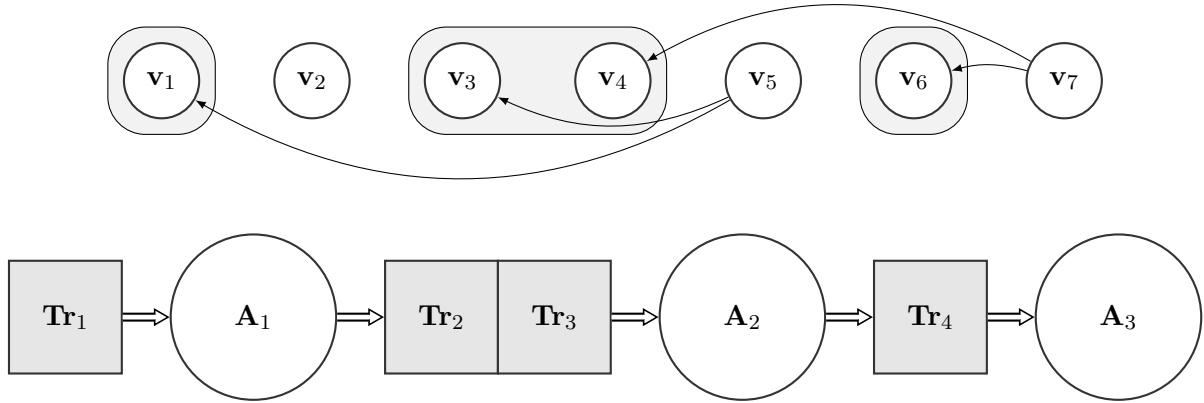
Fig.7  Constellation $H$ with its constellation ordering of vertices (above) and the associated section of the $(c,\lambda,\epsilon)$-$m$-sequence $\chi$ corresponding to $H$.
This section of the sequence $\chi$ consists of three transitive sets and three linear sets.
The second transitive set is partitioned into two equal-length subsets.
Centers of stars as well as singletons ($v_{2},v_{5},v_{7}$) are being looked for
in the linear sets ($A_{1},A_{2},A_{3}$). Leaves of stars are being looked for  in the transitive sets $Tr_{i}$.
\end{center}


We cannot get a strong $(c,\lambda,w)$-$m$-sequence immediately.
It can be constructed if the first approach that we used to find a copy of H or a polynomial-size transitive
subtournament fails.
If at any stage of our analysis in the second part of the algorithm (when we operate on the strong $m$-sequence) 
some star cannot be constructed then we get a linear set complete to/from a big transitive chunk. But now that gives 
us polynomial-size transitive subtournament and the proof is completed.
Note that even though we significantly relax the condition we put on the configurations of stars by introducing
the family of constellations, we still cannot look for a center of the star and some of its leaves in the
same transitive chunk. This is the case since obviously no backward edges can be found in a transitive chunk.
Therefore there are still some limitations put on the configuration of stars. However, as mentioned before, 
there are much weaker than previously. \\
We are ready now to go into technical details of our algorithm.

\subsection{Algorithm~\textit{PolyTrans}}

We start with the \textit{PolyTrans} algorithm that takes as an input an $\mathcal{I}$-strong $N$-big $(c,\lambda,\epsilon)$-$m$-sequence of the $H$-free tournament $T$ and outputs a transitive subtournament of $T$ of size at least $|T|^{\epsilon}$, where $\epsilon=\frac{\log(1-\hat{c})}{\log(\hat{c})}$ for $\hat{c}=\frac{c}{2^{7h^{2}}}$, as long as $\lambda \leq \frac{1}{2^{25h^{2}}h}$, $n=|T| \geq \frac{2^{21h^{2}}}{c}$ and $N \geq 2^{21h^{2}}$.
What is also given as a part of the input for  \textit{PolyTrans} is the procedure $\mathcal{P}$ which, for every subtournament $T_{S}$ of $T$ other than $T$, computes a transitive subtournament of $T_{S}$ of order at least $|T_{S}|^{\epsilon}$.
Assume that $V(H)=\{1,2,...,h\}$ and that $(1,2,...,h)$ is a constellation ordering of $H$. 

\begin{algorithm}
\label{transitivesub} (Algorithm \textit{PolyTrans} returning a transitive subtournament of a polynomial-size)\\
\begin{itemize}
\item \textbf{Input:} An $\mathcal{I}$-strong $(c,\lambda,\epsilon)$-$m$-sequence of length $k=2h^{2}+4h+1$ for $\mathcal{I}=\{h+1,2h+1,...,h^{2}+1\}$ in an $H$-free tournament $T$ of order 
$n \geq \frac{2^{21h^{2}}}{c}$. 
The sequence is $N$-big for $N \geq 2^{21h^{2}}$. Tournament $H$ is a constellation. 
It is assumed that: $\lambda \leq \frac{1}{2^{25h^{2}}h}$ and $\epsilon = \frac{\log(1-\hat{c})}{\log(\hat{c})}$ 
for $\hat{c}=\frac{c}{2^{7h^{2}}}$.
Procedure $\mathcal{P}$ is given that for every subtournament $T_{S}$ of $T$ different than $T$ computes a transitive subtournament 
of $T_{S}$ of order at least $|T_{S}|^{\epsilon}$.
It is assumed that this procedure runs in time $h(|T_{S}|)$ for some given function $h$.
\item \textbf{Output:} A transitive subtournament of $T$ of size at least $|T|^{\epsilon}$.
\item \textbf{Description:} 
Initialize $\sigma$ to be the set of all the stars of $H$.
We first run algorithm~\textit{MakeSmooth} (with $f=0$) on the $m$-sequence from the input  to make the sequence smooth.\\ 
Algorithm~\textit{MakeSmooth} outputs a smooth $(\frac{c}{2},4\lambda, \epsilon)$-$m$-sequence which we denote as: $(C_{1},...,C_{k})$ 
(see: description of \textit{MakeSmooth}). 
We give colors to the vertices of $(C_{1},...,C_{k})$ as follows:
\begin{itemize}
\item if $C_{i}=C_{\zeta(v)}$ for some $v \in V(H)$ and $i$ is even, i.e. $C_{i}$ is transitive, then color the first $\frac{|C_{i}|}{3}$ vertices of $C_{i}$ in the transitive ordering by $v$ and the rest by $h+1$,
\item if $C_{i}=C_{\zeta(v)}$ for some $v \in V(H)$ and $i$ is odd then color an arbitrary subset of $\frac{|C_{i}|}{3}$ vertices of $C_{i}$ by $v$ and the rest by $h+1$,
\item otherwise color all vertices of $C_{i}$ by $h+1$.
\end{itemize}
Then run algorithm~\textit{PolyTransCore} (see: below) on the colored smooth $(\frac{c}{2},4\lambda, \epsilon)$-$m$-sequence described above for parameter $\xi=\frac{1}{6}$, procedure $\mathcal{P}$ and a set of stars $\sigma$.
\end{itemize}
\end{algorithm}

\begin{algorithm}
\label{transitivesubsub} (Algorithm~\textit{PolyTransCore})\\
\begin{itemize}
\item \textbf{Input:} A $(c,\lambda,\epsilon)$-$m$-sequence ($C_{1}$,...,$C_{k}$) of the $H$-free tournament $T$, where vertices of $C_{i}$s are colored by the set $\{1,2,...,h+1\}$, a set of stars $\sigma$ and parameter $\xi>0$. 
It is assumed that if there exists a vertex of $C_{i}$ colored by some color $j$ then at least $\xi |C_{i}|$ vertices of $C_{i}$ are colored by that color.
Procedure $\mathcal{P}$ is given that for every subtournament $T_{S}$ of $T$ different than $T$ computes a transitive subtournament 
of $T_{S}$ of order at least $|T_{S}|^{\epsilon}$.
\item \textbf{Output:} A transitive subtournament of $T$ of size at least $|T|^{\epsilon}$.
\item \textbf{Description:} Take a star $\Sigma^{*} \in \sigma$. Without loss of generality assume that it is a left star. Denote by $c$ its center and by $y_{1},...,y_{q}$ its leaves. Let $X$ be the set of vertices of $C_{\zeta(c)}$ that are colored by $c$ and let $\mathcal{T}^{i}$ be the set of vertices of $C_{\zeta(y_{i})}$ that are colored by $y_{i}$ for $i=1,2,...,q$.
Assume first that there does not exist a vertex $\tau \in X$ and vertices $r_{1},...,r_{q}$ such that $r_{i} \in \mathcal{T}^{i}$ and $\tau$ is adjacent from all $r_{1},...,r_{q}$. Then, by the Pigeonhole Principle, at least $\frac{1}{q}|X|$ vertices of $X$, call this set $\mathcal{C}$, are adjacent to all vertices of $\mathcal{T}^{j^{*}}$ for some $j^{*} \in \{1,2,...,q\}$.
Then run procedure $\mathcal{P}$ on $T|\mathcal{C}$ and merge the transitive subtournament output by the procedure with $\mathcal{T}^{j^{*}}$ to get a bigger transitive subtournament $T^{b}$. Output $T^{b}$.
Assume now that there exists a vertex $\tau \in X$ and vertices $r_{1},...,r_{q}$ such that $r_{i} \in \mathcal{T}^{i}$ for $i=1,2,...,q$ and $\tau$ is adjacent from all $r_{1},...,r_{q}$. Let $D_{i,j}$ be a subset of vertices from $C_{i}$, other than $X,\mathcal{T}^{1},...,\mathcal{T}^{q}$, colored by some fixed color $j$. Denote by $N^{D_{i,j}}_{0} \subseteq D_{i,j}$ the subset of $D_{i,j}$ consisting of vertices of $D_{i,j}$ adjacent from $\tau$ if $\tau$ is before $D_{i,j}$ in the $m$-sequence and adjacent to $\tau$ otherwise (the ordering in the $m$-sequence is induced by an ordering of the sets $C_{i}$ and a transitive ordering within transitive parts of the $m$-sequence). 
Similarly, denote by $N^{D_{i,j}}_{k} \subseteq D_{i,j}$ for $k=1,2,...,q$ the subset of $D_{i,j}$ consisting of vertices of $D_{i,j}$ adjacent from $r_{i}$ if $r_{i}$ is before $D_{i,j}$ in the $m$-sequence and adjacent to $r_{i}$ otherwise. 
Denote $N^{D_{i,j}} = \hat{N}^{D_{i,j}} \bigcap N^{D_{i,j}}_{0}$, where $\hat{N}^{D_{i,j}}=\bigcap_{k=1,2,...,q} N^{D_{i,j}}_{k}$. 
We run algorithm \textit{MakeSmooth} with parameter $f$ satisfying: $(1-f)=\xi(1-\frac{\lambda h}{\xi})$ on the given $m$-sequence $(C_{1},...,C_{k})$, where we take: $S^{j}_{i}=N^{D_{i,j}}$. We obtain new smooth $m$-sequence that we denote as $\chi$. In this new $m$-sequence the coloring is inherited from the old one. We delete $\Sigma^{*}$ from $\sigma$. 
Now we rerun algorithm~\textit{PolyTransCore} on $\chi$ with the updated set of stars and updated parameters $c,\lambda, \xi$: $c \rightarrow \frac{c\xi}{2}(1-\frac{\lambda h}{\xi})$, $\lambda \rightarrow \frac{4\lambda k}{\xi^{2}(1-\frac{\lambda h}{\xi})^{2}}$, $\xi \rightarrow \frac{\xi}{2}(1-\frac{\lambda h}{\xi})$. 
\end{itemize}
\end{algorithm}

The naive implementation of \textit{PolyTrans} algorithm runs in time $h((1-c)n) + \textit{poly}(n)$, where $\textit{poly}(n)$ is an expression polynomial in $n$. This is true since whenever procedure $\mathcal{P}$ is called it is run on a tournament of at most $(1-c)n$ vertices. The correctness of the algorithm~\textit{PolyTrans} is proven later. \\

\subsection{Algorithm \textit{Color-H-free}}

We are ready to give our main algorithm that for a given constellation $H$ with $|H|=h$, properly colors every $H$-free tournament $T$ using at most $|T|^{1-\frac{1}{2^{2^{50h^{2}+1}}}}\log(|T|)$ colors and runs in
time $e^{O(\log(n)^{2})}$. As a byproduct, the algorithm finds in $T$ a transitive subtournament of order at least at least $|T|^{\frac{1}{2^{2^{50h^{2}+1}}}}$.
This tournament is the first color class constructed by the algorithm on the way to produce the entire coloring.

\begin{algorithm}
\label{constellationalgorithm} (Algorithm \textit{Color-H-free} coloring $H$-free tournaments, where $H$ is a constellation) \\
\begin{itemize}
\item \textbf{Input:} Constellation $H$ and $H$-free tournament $T$.
\item \textbf{Output:} Proper coloring of $T$ that uses at most $|T|^{1-\frac{1}{2^{2^{50h^{2}+1}}}}\log(|T|)$ colors.
\item \textbf{Description:} Run algorithm~\textit{Find-L-Sequence} with $\lambda=\frac{1}{2^{2^{9h}}}$ and $k=2h+3$ to get a $l$-sequence $\mathcal{L}$.
Truncate the constructed $l$-sequence $\mathcal{L}$ by deleting its last element. Note that what we get is another $l$-sequence $\mathcal{L}^{t}$. Give it as an input to algorithm~\textit{Find-M-Sequence}. 
As a procedure $\textit{Sub}$ needed by \textit{Find-M-Sequence}(see the description of the algorithm in the next subsection) use algorithm~\textit{Color-H-free} itself 
(thus we use a recursive call of the algorithm~\textit{Color-H-free} on the smaller graph). Denote the output $m$-sequence by $\mathcal{M}$. Truncate it (by deleting its last elements) to reduce its length to exactly $2^{h+2}(h+1)+2h+1$. Denote the truncated $m$-sequence as $\mathcal{M}^{t}$. Now run on $\mathcal{M}^{t}$ algorithm~\textit{FindStrong-M-Sequence} 
to get an $\mathcal{I}$-strong $m$-sequence $\mathcal{M}^{s}$. Give $\mathcal{M}^{s}$ as an input to algorithm~\textit{PolyTrans}. Again, as a procedure $\mathcal{P}$ needed by \textit{FindStrong-M-Sequence} use algorithm~\textit{Color-H-free} itself. 
As an output we obtain a transitive subtournament $T_{1}$. 
This transitive tournament becomes a first element of the output of the algorithm \textit{Color-H-free} (remember that \textit{Color-H-free} outputs both: polynomial-size transitive subset and the coloring). The remaining element of the output is the coloring and we are about to give it.
Color all vertices of $T_{1}$ by $1$. Remove $T_{1}$ from $T$ and repeat the entire procedure to obtain a transitive subtournament $T_{2}$. Color all its vertices by $2$. Remove $T_{2}$ from $T$ and keep repeating the procedure. Continue until the tournament you are left with is empty. By that time all the vertices of $T$ were colored and one can easily note that this is a proper coloring of $T$. The color classes are the sets: $V(T_{1})$, $V({T_{2}})$,... 
Output that coloring.
\end{itemize}
\end{algorithm}

\subsection{Technical algorithms}

\subsubsection{Introduction}

In this section we present algorithms: \textit{Find-L-Sequence}, \textit{Find-Clique}, \textit{Find-M-Sequence}, \textit{MakeSmooth}, \textit{FindStrong-M-Sequence}
that serve as technical subroutines for algorithms introduced earlier.

\subsubsection{Algorithm \textit{Find-L-Sequence}}

For a given $0 < \lambda < 1$ and $k \geq 0$ let $\lambda_{i}=(\frac{\lambda^{2}}{4^{k}h^{4k}})^{2^{i}h^{2i}}$ for $i=0,1,...,k$.

We start by describing \textit{Find-L-Sequence} algorithm that calculates a $(c,\lambda)$-$l$-sequence of length $2^{k}$ for some given $k \geq 0$ in 
the $H$-free tournament $T$ with: $|T| \geq \frac{2^{k+1}(h+1)h^{2k}}{\lambda_{k}^{hk}}$, where $H$ is an arbitrary tournament with $|H| \geq 2$, $V(H)=\{v_{1},...,v_{h}\}$, and $c=c(H,k,\lambda)=\frac{\lambda_{k}^{|H|k}}{2^{k}|H|^{2k}}$.
The algorithm \textit{Find-L-Sequence} uses subroutine~\textit{Find-L-Sequence-Core} which recursively runs algorithm~\textit{Find-L-Sequence} on the smaller input.

\begin{algorithm}
\label{lsequencealgol}
(Algorithm \textit{Find-L-Sequence} constructing a $(c,\lambda)$-$l$-sequence in an $H$-free tournament) \\
\begin{itemize}
\item \textbf{Input}: $(\lambda,k, H, T)$, where $T$ is $H$-free, $|H| \geq 2$. 
\item \textbf{Output}: A $(c,\lambda)$-$l$-sequence of length $2^{k}$ in $H$, where $c=c(H,k,\lambda)=\frac{\lambda_{k}^{hk}}{2^{k}h^{2k}}$.
\item \textbf{Description}: If $k=0$ then output $V(T)$. Assume now that $k>0$. Then choose in $T$ arbitrarily $h$ pairwise disjoint sets: $S_{1},...,S_{h}$, 
each of size $ \lfloor  \frac{n}{h+1} \rfloor$, where $n=|T|$. Run subroutine~\textit{Find-L-Sequence-Core} with parameters: $(H,\lambda,\lambda_{k}, k, H, T|S_{1} \cup ... \cup S_{h}, S_{1},...,S_{h})$, where $h=|H|$.
\end{itemize}
\end{algorithm}

\begin{subroutine}
\label{lsequence2algol}
(Subroutine \textit{Find-L-Sequence-Core})\\
\begin{itemize}
\item \textit{\textbf{Input}}: $(H,\lambda,\lambda_{k}, k, H^{r}, T^{r}, S_{t_{1}},...,S_{t_{|H^{r}|}})$, where $k>0$, $|H|>1$, $V(H^{r})=\{v_{t_{1}},...,v_{t_{|H^{r}|}}\}$ for some vertices: $v_{t_{1}},...,v_{t_{|H^{r}|}}$ and $T^{r}$ is $H$-free. 
It is assumed that: $S_{t_{1}},...,S_{t_{|H|^{r}}}$ are nonempty.
\item \textit{\textbf{Output}}: Sequence of pairwise disjoint subsets of $V(T)$ of length $2^{k}$ (see: Description below).
\item \textit{\textbf{Description}}: If $|H^{r}|=1$ then raise an exception (in the analysis of the algorithm we will show that the exception in fact will never be raised). Now assume that $|H^{r}|>1$. Denote $h^{r}=|H^{r}|$. For a vertex $v \in S_{t_{1}}$ and an index $j=2,3,...,h^{r}$ denote by $N_{j}(v)$ the set of vertices of $S_{t_{j}}$ adjacent from $v$ if $v_{t_{j}}$ is adjacent from $v_{t_{1}}$ or adjacent to $v$ if $v_{t_{j}}$ is adjacent to $v_{t_{1}}$ in $H^{r}$. Calculate values $|N_{j}(v)|$ for every $v \in S_{t_{1}}$ and $j=2,3,...,h$. 

If there exists a vertex $v^{*} \in S_{t_{1}}$ such that $|N_{j}(v^{*})| \geq \lambda_{k} |S_{t_{j}}|$ for every $j \in \{2,3,...,h^{r}\}$ then run recursively subroutine~\textit{Find-L-Sequence-Core} with parameters: \\
$(H,\lambda,\lambda_{k},k,H^{r} \backslash \{v_{t_{1}}\}, T^{r}|N_{2}(v^{*}) \cup ... \cup N_{h^{r}}(v^{*}),N_{2}(v^{*}),...,N_{h^{r}}(v^{*}))$. 

If this is not the case then (by Pigeonhole Principle) at least $\frac{|S_{t_{1}}|}{h^{r}-1}$ vertices $v$ of $S_{t_{1}}$ satisfy the following: there exists $j^{*} \in \{2,3,...,h^{r}\}$ such that $|N_{j^{*}}(v)| < \lambda_{k} |S_{t_{j^{*}}}|$. Denote this set of vertices by $W$. Run algorithm~\textit{Find-L-Sequence} with parameters: $(\lambda, k-1, H, T^{r}|W)$ to get a $l$-sequence: $(A_{1},...,A_{2^{k-1}})$.
Run algorithm~\textit{Find-L-Sequence} with parameters: $(\lambda, k-1, H, T^{r}|S_{t_{j^{*}}})$ to get a $l$-sequence: $(A^{'}_{1},...,A^{'}_{2^{k-1}})$.
Note that $d(W,S_{t_{j^{*}}}) \geq 1-\lambda_{k}$ or\\ $d(S_{t_{j^{*}}},W) \geq 1-\lambda_{k}$. If the former is true then output the sequence $(A_{1},...,A_{2^{k-1}},A^{'}_{1},...,A^{'}_{2^{k-1}})$. 
Otherwise output the sequence $(A^{'}_{1},...,A^{'}_{2^{k-1}},A_{1},...,A_{2^{k-1}})$.  
\end{itemize}
\end{subroutine}

Algorithm~\textit{Find-L-Sequence} runs in polynomial time. Its correctness will be proven later.
Later we will also analyze its running time in more detail.

\subsubsection{Algorithm \textit{Find-Clique}}

For an undirected graph $G$ and two nonempty disjoint sets: $A,B \subseteq V(G)$ we define, by an analogy to the directed setting: $d(A,B) = \frac{e_{A,B}}{|A||B|}$, where $e_{A,B}$ is the number of edges between $A$ and $B$.
Assume now that we have a $k$-partite undirected graph with color classes:
$V_{1},...,V_{k}$. Assume besides that for every $1 \leq i < j \leq k$ the following holds: $d(V_{i},V_{j}) \geq 1 - \lambda$, where: $\lambda \leq \frac{1}{3^{2k+1}k}$.
Now we will present an algorithm \textit{Find-Clique} that finds in this graph a clique: $\{v_{1},...,v_{k}\}$ such that $v_{i} \in V_{i}$ for $i=1,2,...,k$.

\begin{algorithm}
\label{cliquealgorithm}
(Algorithm \textit{Find-Clique} finding a clique in a dense k-partite undirected graph)
\begin{itemize}
\item \textbf{Input:} $k$-partite undirected graph with color classes: $V_{1},...,V_{k}$ such that for every $1 \leq i < j \leq k$ the following holds: $d(V_{i},V_{j}) \geq 1 - \lambda$, where $\lambda$ satisfies: $0 < \lambda < \frac{1}{3^{2k+1}k}$.
\item \textbf{Output:} A clique $\{v_{1},...,v_{k}\}$ such that $v_{i} \in V_{i}$ for $i=1,2,...,k$. 
\item \textbf{Description:} If $k=1$ then output an arbitrary vertex of $V_{1}$.
Now assume that $k>1$. Define $W_{i}=\{v \in V_{1}:d(\{v\},V_{i})<1-2k\lambda\}$ for $i=2,3,...,k$. For every $i$ each set $W_{i}$ may be easily computed. Having sets $W_{i}$, take a vertex $v_{1} \in V_{1} \backslash (W_{2} \cup ... \cup W_{k})$. For every $i=2,3,...,k$ compute $V^{'}_{i}=V_{i} \cap N^{v}_{i}$, where $N^{v}_{i}$ is the set of vertices adjacent to $v_{1}$ in $V_{i}$. Run recursively algorithm~\textit{Find-Clique} on the $(k-1)$-partite induced subgraph with color classes: $V^{'}_{2},...,V^{'}_{k}$,
and update $\lambda$:  $\lambda \rightarrow \frac{\lambda}{(1-2k\lambda)^{2}}$.
You obtain a set $\{v_{2},...,v_{k}\}$. Output the set $\{v_{1},...,v_{k}\}$.
\end{itemize}
\end{algorithm}

The algorithm clearly runs in $\textit{poly}(\max_{i=1,...,k}|V_{i}|)$ time (note that $k$ is a constant). Its correctness will proven later.

\subsubsection{Algorithm \textit{Find-M-Sequence}}

Let $T$ be a tournament with $n$ vertices. Assume now that $(A_{1},...,A_{2k+1})$ is a $(c,\lambda)$-$l$-sequence in $T$, where $\lambda \leq \frac{\Lambda}{4(2k+1)3^{4k+3}}$. Assume furthermore than there exists a subroutine \textit{Sub} that for every $i=1,2,...,k$ and for every subtournament $T_{S}$ of $A_{2i}$ computes in time $O(h(|T_{S}|))$ a transitive subtournament of $T_{S}$ of size at least $|T_{S}|^{\epsilon}$ for some given $\epsilon>0$.
Under these conditions we will show an algorithm \textit{Find-M-Sequence} that computes in $T$
a $(c^{'},\Lambda,\epsilon)$-$m$-sequence of length $2k+1$
for $c^{'}=\min((\frac{c}{2})^{\epsilon},c)$ which in addition is $(\log(cn)-2)$-big. 
From the characteristic of the sequence we already know that the sequence is $c^{'}n$-big and
asymptotically for large $n$ this is a stronger property than being $(\log(cn)-2)$-big. However for
the corner cases for small $n$ (that we will need for induction to prove that the algorithm produces a transitive subset in the $H$-free tournament of the desired size) we will also need $(\log(cn)-2)$-bigness property.

\begin{algorithm}
\label{msequencealgol} (Algorithm \textit{Find-M-Sequence} finding $m$-sequences) \\
\begin{itemize}
\item \textbf{Input:} A $(c,\lambda)$-$l$-sequence $(A_{1},...,A_{2k+1})$ in $T$ for $\lambda \leq \frac{\Lambda}{4(2k+1)3^{4k+3}}$ for some parameter $0 < \Lambda < 1$ and a subroutine \textit{Sub} that for every $i=1,2,...,k$ and for every subtournament $T_{S}$ of $A_{2i}$ computes in time $O(h(|T_{S}|))$ a transitive subtournament of $T_{S}$ of size at least $|T_{S}|^{\epsilon}$ for some given $\epsilon>0$. 
\item \textbf{Output:} A $(\min((\frac{c}{2})^{\epsilon},c),\Lambda,\epsilon)$-$m$-sequence of length $2k+1$ in $T$ which is $(\log(cn)-2)$-big.
\item \textbf{Description:} Denote $n=|T|$. Let $i \in {1,2,...,k}$. 
Compute a transitive subset $Tr^{i}_{1} \subseteq A_{2i}$ of size at least $\log(\frac{cn}{2})-1$ for $i=1,2,...,k$. Such a subset always exists since $|A_{2i}| \geq \frac{cn}{2}$ (it also can be efficiently computed, we will discuss this in more detail later). Now use subroutine \textit{Sub} to compute a transitive subset $Tr^{i}_{2} \subseteq A_{2i}$ of size $\lceil (\frac{c}{2})^{\epsilon} n^{\epsilon} \rceil$.
Remove from $A_{2i}$ the bigger subset from the set: $\{Tr^{i}_{1},Tr^{i}_{2}\}$ and denote this bigger one by $T^{i}_{1}$. Continue until the set you are left with is of size smaller than $\frac{|A_{2i}|}{2}$. Denote transitive subsets obtained in such a way as: $T^{i}_{1},...,T^{i}_{r_{i}}$. Construct a $(2k+1)$-partite graph $G$ with color classes: $V_{1},...,V_{2k+1}$ as follows:
\begin{itemize}
\item $V_{2i+1}=\{A_{2i+1}\}$ for $i=1,2,...,k$,
\item $V_{2i}=\{T^{i}_{1},...,T^{i}_{r_{i}}\}$ for $i=1,2,...,k$, 
\item make two vertices $X \in V_{a}$, $Y \in V_{b}$ for $1 \leq a < b \leq 2k+1$ adjacent if $d(X,Y) \geq 1 - \Lambda$. 
\end{itemize}
Run algorithm~\textit{Find-Clique} on $G$ to obtain a sequence $(Y_{1},...,Y_{2k+1})$ that induces a clique in $G$ ($Y_{i} \in V_{i}$ for $i=1,2,...,2k+1$). Output $(Y_{1},...,Y_{2k+1})$. 
\end{itemize}
\end{algorithm}

Denote $A^{m}=\max_{i=1,...,2k+1}|A_{i}|$.
The algorithm runs in time $O(h(A^{m}))n^{1-\epsilon} +\textit{poly}(n)$, where $\textit{poly}(n)$ is a polynomial factor. Its correctness and running time will be proven later.

\subsubsection{Algorithm \textit{MakeSmooth}}

Assume that we are given a sequence $(C_{1},...,C_{k})$ that is either a $(c,\lambda,\epsilon)$-$m$-sequence, a $(c,\lambda,\epsilon)$-$t$-sequence or an $(c,\lambda)$-$l$-sequence. 
Assume that $S^{j}_{i} \subseteq C_{i}$ for $j=1,2,...,s_{i}$. Assume furthermore that $S^{j}_{i}$s are pairwise disjoint for $j=1,2,...,s_{i}$ and $|S^{j}_{i}| \geq (1-f)|C_{i}|$ for $i=1,2,...,k$, $j=1,2,...,s_{i}$, where $0 < f < 1$ is some fixed parameter. Denote $L=s_{1}+...+s_{k}$.
We will show an algorithm that extracts a subset from every $S^{j}_{i}$ and uses extracted subsets to construct a smooth $(c^{'},\lambda^{'},\epsilon)$-$m$-sequence, a smooth $(c^{'},\lambda^{'},\epsilon)$-$t$-sequence or a smooth $(c^{'},\lambda^{'})$-$l$-sequence respectively, where $c^{'}=\frac{c}{2}(1-f)$ and $\lambda^{'}=\frac{4\lambda L}{(1-f)^{2}}$.
Each element of the constructed sequence ($m$-sequence, $t$-sequence or $l$-sequence)
is the union of the corresponding extracted subsets. The goal is to make an input sequence smooth by
taking subsets, but in such a way that a significant fraction of elements from each part of the partition defined by sequences $S^{j}_{i}$ is used in the built smooth sequence.

The naive implementation of the algorithm below clearly runs in polynomial time. Its correctness is proven in the next section.\\

\begin{algorithm}
\label{smoothingalgorithm} (Algorithm \textit{MakeSmooth})\\
\begin{itemize}
\item \textbf{Input:} A $(c,\lambda,\epsilon)$-$m$-sequence, a $(c,\lambda,\epsilon)$-$t$-sequence or 
a $(c,\lambda)$-$l$-sequence $(C_{1},...,C_{k})$, a set of subsets $S^{j}_{i}$ for $i=1,2,...,k$, $j=1,2,...,s_{i}$ 
such that $S^{j}_{i} \subseteq C_{i}$, $|S^{j}_{i}| \geq (1-f)|C_{i}|$ and $S^{j}_{i}$ are pairwise disjoint for $i=1,2,...,k$, $j=1,2,...,s_{i}$ and some $0<f<1$. It is assumed that $s_{i}=1$ for $i=1,2,...,k$ if $(C_{1},...,C_{k})$ is a $l$-sequence, and $s_{i}=1$ for $i=1,3,5,...$ if $(C_{1},...,C_{k})$ is a $m$-sequence (in other words, if $s_{i}>1$ then $i$ corresponds to the transitive set in $(C_{1},...,C_{k})$).
Denote $L=s_{1}+...+s_{k}$. 
\item \textbf{Output:}  A smooth $(c^{'},\lambda^{'},\epsilon)$-$m$-sequence, a smooth $(c^{'},\lambda^{'},\epsilon)$-$t$-sequence or a smooth $(c^{'},\lambda^{'})$-$l$-sequence $(C^{'}_{1},...,C^{'}_{k})$ respectively, where $c^{'}=\frac{c}{2}(1-f)$, $\lambda^{'}=\frac{4\lambda L}{(1-f)^{2}}$ and $C^{'}_{i} \subseteq \bigcup_{j=1,2,...,s_{i}} S^{j}_{i}$ for $i=1,2,...,k$.  Besides we have: $|C^{'}_{i} \cap S^{j}_{i}| \geq \frac{cn}{2}(1-f)$ for $i=1,2,...,k$, $j=1,2,...,s_{i}$.
\item \textbf{Description:} Let us assume that $(C_{1},...,C_{k})$ is a $(c,\lambda,\epsilon)$-$m$-sequence. For two remaining cases the algorithm is completely analogous. 
For $i,j \in \{1,2,...,k\}$, $i \neq j$, $t_{1} \in \{1,2,...,s_{i}\}$, $t_{2} \in \{1,2,...,s_{j}\}$ denote by $C^{j,t_{2}}_{i,t_{1}} \subseteq S^{t_{1}}_{i}$ the set of vertices of $S^{t_{1}}_{i}$ that: 
\begin{itemize}
\item are adjacent to at least $(1-\frac{2L\lambda}{(1-f)^{2}})|S^{t_{2}}_{j}|$ vertices of $S^{t_{2}}_{j}$ if $i<j$ or
\item are adjacent from at least $(1-\frac{2L\lambda}{(1-f)^{2}})|S^{t_{2}}_{j}|$ vertices of $S^{t_{2}}_{j}$ if $i>j$.
\end{itemize}
Take $C^{i,t_{1}} = \bigcap_{j \neq i, t_{2}=1,2,...,s_{j}} C^{j,t_{2}}_{i,t_{1}}$ for $i=1,2,...,k$, $t_{1}=1,2,...,s_{i}$. Denote $C^{'}_{i}=\bigcup_{t=1,2,...,s_{i}} C^{i,t}$ for  $i=1,2,...,k$ and output $(C^{'}_{1},...,C^{'}_{k})$. 
\end{itemize}
\end{algorithm}

\subsubsection{Algorithm \textit{FindStrong-M-Sequence}}

We will now show a technical algorithm that is fundamental for finding efficient coloring of an $H$-free tournament, where $H$ is a constellation.
Again as before, the naive implementation of the algorithm runs in a polynomial time. Its correctness is proven in the next section.
The presented procedure is essentially a wrapper for the main algorithm \textit{FindStrong-M-Sequence-Main} described next to it. This main algorithm operates on the smooth $m$-sequence
that was obtained in the preprocessing performed in the initial phase of \textit{FindStrong-M-Sequence}.
The goal of the  \textit{FindStrong-M-Sequence} is to extract a strong $m$-sequence from an $H$-free tournament, where $H$ is a given constellation. The input to the procedure is a $M$-big $m$-sequence
for an appropriate parameter $M$.

Let us remind now some important terms regarding constellations.
The \textit{interstellar graph} of the constellation $H$ under a star ordering $\theta$ is an undirected graph, whose vertices are the sets of leaves of the stars of $H$ under $\theta$ and any two given vertices $L_{1}$ and $L_{2}$ are adjacent iff:
\begin{itemize}
\item the left point of $L_{1}$ precedes the right point of $L_{2}$ in $\theta$ and
\item the left point of $L_{2}$ precedes the right point of $L_{1}$ in $\theta$.
\end{itemize}  
Take an interstellat graph of the constellation $H$.
For each connected component $C$ of the interstellar graph of $H$ we denote by $Z(C)$ the union of subsets of $V(H)$ corresponding to its vertices (this is the union of some subsets of $V(H)$ of the vertices of $H$). 
We define $\mathcal{C}({Z(C)})$ as follows.
We say that a vertex $v \in \mathcal{C}({Z(C)})$ if $v \in Z(C)$ or $v$ is between some two vertices of $Z(C)$ under the ordering $\theta$.
Let $C_{1},...,C_{k}$ be the connected components of the interstellar graph. Note that for any given $1 \leq i < j \leq k$ either every vertex of $\mathcal{C}({Z(C_{i})})$ is before every vertex of $\mathcal{C}({Z(C_{j})})$, or every vertex of $\mathcal{C}({Z(C_{j})})$ is before every vertex of $\mathcal{C}({Z(C_{i})})$. Thus there is a natural ordering of the sets $\mathcal{C}({Z(C_{i})})$ for $i=1,2,...,k$ induced by the ordering of the vertices. Denote the ordered sequence of the sets $\mathcal{C}({Z(C_{i})})$ for $i=1,2,...,k$  as 
$(\mathcal{W}_{1},...,,\mathcal{W}_{k})$, where a set $\mathcal{W}_{i}$ is before a set $\mathcal{W}_{j}$ for $1 \leq i < j \leq k$. 
Denote $\mathcal{W}_{0}=\mathcal{W}_{k+1} = \emptyset$.
For $i=1,2,...,k+1$ denote by $\mathcal{R}_{i}$  the set of the vertices of $H$ that are after all the vertices of $\mathcal{W}_{i-1}$ and before all the vertices of $\mathcal{W}_{i}$ under the ordering $\theta$.  
Note that if $\mathcal{R}_{i}$ is nonempty then all its elements are centers of the stars of $H$.
Denote the set of nonempty sets $\mathcal{R}_{i}$ as $\{\mathcal{M}_{1},...,\mathcal{M}_{r}\}$ for some $r \geq 0$.
Note that $\{\mathcal{W}_{1},...,\mathcal{W}_{k},\mathcal{M}_{1},...,\mathcal{M}_{r}\}$  is a partition of the vertices of $H$.
We denote this partition by $P_{\theta}(H)$.
Each constellation satisfies the following: if some center of the star belongs to some $P \in P_{\theta}(H)$ then no leaf of this star belongs to $P$.

Let us explain what the algorithm \textit{FindStrong-M-Sequence-Main} is doing. We commented on it before when we were talking about techniques used in the algorithm, but now we will be more precise. The input to the algorithm
is a long $m$-sequence. The algorithm tries to reconstruct a constellation $H$ in the given tournament.
It wants to achieve it by selecting $|P_{\theta}(H)|$ transitive chunks, where $\theta$ is a constellation ordering of $H$, mapping each vertex of the constellation to one of the selected $|P_{\theta}(H)|$
chunks and looking for it in the chunk it was mapped to (if several vertices are mapped to the same chunk then the algorithm tries to find them in different subchunks in the subdivision of the given chunk).
Each transitive element of the given $m$-sequence is a vertex of the undirected graph $G$ that encodes the relation between different transitive chunks. An edge between two nodes in $G$ indicates that throughout the execution of the algorithm a particular behaviour between corresponding transitive subtournaments was detected, namely one was detected to be adjacent to the other one. This type of relation is particularly precious since both transitive chunks were previously extracted from different linear sets so there was no reason to assume before that the relation was true (it will ultimately enable us to look for different leaves of the same star in different transitive elements of the $m$-sequence when we will look for $H$ later using different approach). Graph $G$ evolves during the execution of the algorithm as new pairs of transitive chunks such that one is adjacent to the other one are detected (the evolution is conducted by adding new edges as well as replacing with new transitive chunks the old ones in the vertex set  $V(G)$). The algorithm tries to reconstruct $H$ in the tournament induced by selected $|P_{\theta}(H)|$ transitive chunks star by star. Whenever vertices inducing a particular star are found we say that \textit{state 1 was reached}.
Since the reconstruction of the entire $H$ cannot succeed (input tournament is $H$-free) at some point, by simple analysis based on the Pigeonhole Principle, the algorithm detects two substantial transitive subchunks, such that one is adjacent to the other one. Since the initial $|P_{\theta}(H)|$ transitive chunks were 
chosen as an independent set in $G$, when we replace in $G$ the original two transitive chunks by the two found transitive subchunks, we also need to add one more edge. 
So from the point of view of graph $G$ in each step we are taking its independent set of size $|P_{\theta}(H)|$ and we add an edge between some two vertices of this set.
Since the graph has $2^{h+1}$ nodes (and this number is the same throughout the execution of the algorithm), at every single step it has a clique or an independent set of size $h$. If an independent set of size $h$ does not exist then we take an $h$-clique and it it easy to see that it corresponds to the strong $m$-sequence (every edge of the clique indicates the relation: \textit{adjacent to} between corresponding transitive subsets). Simple calculations lead to the conclusion that the constructed strong $m$-sequence has desired characteristic (in terms of size of its elements, etc). If an independent set is found then new edge is added. Now notice that since whenever there exists an independent set of size $h$ an edge is added, at some point the clique of size $h$ will appear in $G$ anyway. Thus we will be always able to construct a strong sequence we are looking for.

\begin{algorithm}
\label{strongstructure} (Algorithm \textit{FindStrong-M-Sequence} constructing a strong $m$-sequence)\\
\begin{itemize}
\item \textbf{Input:} A constellation $H$ with $V(H)=\{1,2,...,h\}$, a constellation ordering $\theta=(1,2,...,h)$, an $H$-free tournament $T$ and a $(c,\lambda,\epsilon)$-$m$-sequence in $T$ of length $k=2^{h+2}(h+1)+2h+1$ which is $M$-big for $M \geq 2(h+2) \cdot 2^{2^{8h+2}}$. It is assumed that $\lambda \leq \frac{1}{2^{2^{5h+6}}}$.
\item \textbf{Output:} An $\mathcal{I}$-strong $(\hat{c},\hat{\lambda},\epsilon)$-$m$-sequence in $T$ of length $\hat{k}$ which is $\frac{1}{2^{2^{4h+3}}}M$-big, where: $\hat{c}=\frac{1}{2^{2^{4h+3}}}c$, $\hat{\lambda}=2^{2^{5h}} \lambda$, $\hat{k}=2h^{2}+4h+1$, $\mathcal{I}=\{h+1,2h+1,...,h^{2}+1\}$.
\item \textbf{Description:} Run algorithm~\textit{MakeSmooth} to get a smooth $(c^{'},\lambda^{'},\epsilon)$-$m$-sequence $(C_{1},...,C_{k})$, where $c^{'}=\frac{c}{2}$, $\lambda^{'}=4 \lambda k$ (in algorithm~\textit{MakeSmooth} we take $f=0$, $s_{i}=1$ for $i=1,2,...,k$ and $S^{j}_{i}=C_{i}$).
Let $\chi=2h+1$. Denote transitive sets: $C_{\chi+1},C_{2\chi+2},C_{3\chi+3}...,C_{2^{h+1}\chi+2^{h+1}}$ as $T_{1},...,T_{2^{h+1}}$ respectively.
Then run algorithm~\textit{FindStrong-M-Sequence-Main} (see description below), where the arguments for  \textit{FindStrong-M-Sequence-Main} are defined as follows: 

\begin{itemize}
\item input $m$-sequence is a smooth $(c^{'},\lambda^{'},\epsilon)$-$m$-sequence $(C_{1},...,C_{k})$ computed above,
\item input graph $G$ is an undirected graph with $V(G)=\{T_{1},...,T_{2^{h+1}}\}$ and no edges,
\item an independent set $S$ is of the form: $S=\{T_{1},...,T_{h}\}$,
\item input parameter $\sigma$ is the set of all stars of $H$,
\item input parameter $\xi$ satisfies: $\xi=\frac{1}{2(h+2)}$,
\end{itemize}

and the coloring of vertices of the $m$-sequence is done as follows:
\begin{itemize}
\item for every $T_{i}$, $i=1,2,...,h$ color $j$ for $j=1,2,...,h$ is assigned to the vertices of the indices: $(j-1)\lfloor\frac{|T_{i}|}{h+2}\rfloor+1,...,j\lfloor\frac{|T_{i}|}{h+2}\rfloor$ in the transitive ordering of $T_{i}$ (we use the convention that the first vertex in the ordering has index $1$), all other vertices of $T_{i}$ are colored by $h+1$,
\item vertices of all other sets of the $m$-sequence are colored by $h+1$.
\end{itemize}
\end{itemize}
\end{algorithm}

\begin{algorithm}
\label{strongstructuresubroutine} (Algorithm \textit{FindStrong-M-Sequence-Main})\\
\begin{itemize}
\item \textbf{Input:} A constellation $H$ with $V(H)=\{1,2,...,h\}$, a constellation ordering $\theta=(1,2,...,h)$, an $H$-free tournament $T$ and a smooth $(c,\lambda,\epsilon)$-$m$-sequence $(C_{1},...,C_{k})$ in $T$ of length $k=2^{h+2}(h+1)+2h+1$. Every vertex of the $m$-sequence is colored by a color from the set $\{1,2,...,h+1\}$.
For every color $j \in \{1,...,h+1\}$ and every $i \in \{1,...,k\}$ if there are vertices in $C_{i}$ colored by $j$ then at least $\xi |C_{i}|$ of them are colored by $j$. Furthermore, a nonempty subset $\sigma$ of the set of stars of $H$ is given. 
We are also given an undirected graph $G$ with $V(G)=\{C_{\chi+1},C_{2\chi+2},C_{3\chi+3}...,C_{2^{h+1}\chi+2^{h+1}}\}$, where $\chi=2h+1$, and an independent set $S$ of $G$ denoted as $S=\{T^{'}_{t_{1}},...,T^{'}_{t_{h}}\}$ for $t_{1} < t_{2} < ... <t_{h}$.
\item \textbf{Output:} An $\mathcal{I}$-strong $m$-sequence of length $\hat{k}=2h^{2}+4h+1$, where $\mathcal{I}=\{h+1,2h+1,...,h^{2}+1\}$.
\item \textbf{Description:} Take an arbitrary star $\Sigma^{*} \in \sigma$. Take a partitioning $P_{\theta}(H)$ and let assume that it is of the form: $P_{\theta}(H)=\{P_{1},...,P_{z}\}$, where vertices of $P_{i}$ are before vertices of $P_{j}$ under an ordering $\theta$ for $i < j$ and $z$ is the number of elements of the partition $P_{\theta}(H)$  (see: definition of $P_{\theta}(H)$).
We will assume that $\Sigma^{*}$ is a left star. For a right star the algorithm is completely analogous. Let $n_{c}$ be such that the center of $\Sigma^{*}$ is in $P_{n_{c}}$ under $\theta$. Denote by $m_{c}$ the position that this center occupies in $P_{n_{c}}$  under ordering $\theta$ (first vertex of $P_{n_{c}}$ under ordering $\theta$ occupies position $1$, second  - position $2$, etc.). Assume that $\Sigma^{*}$ has $q$ leaves and that all the leaves are in $P_{n_{l}}$ for some $n_{l}$ (note that from the definition of the constellation, $n_{l}$ is the same for all leaves of $\Sigma^{*}$ and is different than $n_{c}$). Denote by $m_{i}$ for $i=1,2,...,q$ the position that $i^{th}$ leaf occupies in $P_{n_{l}}$ under ordering $\theta$. First we check whether there exists a vertex $\rho \in T^{'}_{t_{n_{c}}}$ that is colored by $m_{c}$ and vertices $r_{1},...r_{q}$ such that:
\begin{itemize}
\item $r_{i} \in T^{'}_{t_{n_{l}}}$,
\item $r_{i}$ is colored by $m_{i}$ for $i=1,2,...,q$ and
\item $(r_{i},\rho)$ is a backward edge for $i=1,...,q$ .
\end{itemize}

As we have already noticed, since $H$ is a constellation we know that $n_{c} \neq n_{l}$.
If vertices $\rho, r_{1},...,r_{q}$ exist we say that \textit{state $1$} was reached. Otherwise we say that \textit{state $0$} was reached. 

Assume first that state $0$ was reached. But then, by Pigeonhole Principle, there exists a subset $\mathcal{C} \subseteq T^{'}_{t_{n_{c}}}$ of at least $\frac{1}{q}| T^{'}_{t_{n_{c}}}|$ vertices that are adjacent to all vertices of $T^{'}_{t_{n_{l}}}$ colored by some fixed color $i^{*} \in \{1,2,...,h\}$. 
Indeed, if state $0$ was reached then we could not construct the embedding of the left star defined above. So no matter which vertex $v$ of $T^{'}_{t_{n_{c}}}$ is taken as the center, the construction
is not possible. Fix such a vertex $v$. We try to find leaves of the star in differently colored chunks, i.e. find a backward edge from a colored chunk to $v$ for every color. 
If this is not possible then $v$ is adjacent to all vertices for some particular color $col_{v}$. This is the place where the Pigeonhole Principle comes into action.
Since altogether we have $q$ colors, for at least   $\frac{1}{q}|T^{'}_{t_{n_{c}}}|$ vertices $v$ from  $T^{'}_{t_{n_{c}}}$ the color $col_{v}$ will be the same.
In other words, at least $\frac{1}{q}|T^{'}_{t_{n_{c}}}|$ vertices $v$ from  $T^{'}_{t_{n_{c}}}$ will be adjacent to all vertices of $T^{'}_{t_{n_{l}}}$ colored by some fixed color $i^{*} \in \{1,2,...,h\}$. 

In this scenario we replace $T^{'}_{t_{n_{c}}}$ in the $m$-sequence by $\mathcal{C}$ and $T^{'}_{t_{n_{l}}}$ by the subset $\mathcal{L}$ of $T^{'}_{t_{n_{l}}}$ consisting of vertices colored by $i^{*}$. In the undirected graph $G$ we replace vertex $T^{'}_{t_{n_{c}}}$ by $\mathcal{C}$, vertex $T^{'}_{t_{n_{l}}}$ by $\mathcal{L}$ (keeping all edges of $G$, new vertices inherit edges adjacent to vertices that they replaced) and add an edge between vertex $\mathcal{C}$ and vertex $\mathcal{L}$. Then we run on our updated $m$-sequence (which is not necessarily smooth) an algorithm~\textit{MakeSmooth} to make it smooth (with the same parameters as in the preprocessing phase of the algorithm~\textit{FindStrong-M-Sequence}). In $G$ we replace all vertices by corresponding subsets extracted from them during smoothing-procedure (edges are inherited from the old graph $G$). Then we recolor all the vertices of the new $m$-sequence we obtained using the same coloring procedure that we used earlier in the algorithm~\textit{FindStrong-M-Sequence} before calling algorithm~\textit{FindStrong-M-Sequence-Main} for the first time. We replace our collection of stars by the collection of all stars of $H$ which we call $\sigma_{H}$. 
We check whether there is a clique of size $h$ in $G$. Assume first that there is not. Then, since $G$ has $2^{h+1}$ vertices, it has an independent set $S^{*}$ of size $h$. We rerun algorithm~\textit{FindStrong-M-Sequence-Main} with updated parameters $c,\lambda, \xi, \sigma, S$: $c \rightarrow \frac{c\xi}{2h}$, $\lambda \rightarrow \frac{4\lambda h^{2} k}{\xi^{2}}$, $\xi \rightarrow \frac{1}{2(h+2)}$, $\sigma \rightarrow \sigma_{H}, S \rightarrow S^{*}$ and updated graph $G$.        
Assume now that the clique of size $h$ was found.
Then note that we can easily extract from $(C_{1},...,C_{k})$ an $\mathcal{I}$-strong subsequence of length $\hat{k}$ (this subsequence in particular contains all vertices of the clique). We output it. 

It remains to consider scenario when state $1$ was reached.
If this is the case we remove $\Sigma^{*}$ from $\sigma$. 
Let $X$ be the set of vertices from $T^{'}_{t_{n_{c}}}$ colored by $m_{c}$ and let $Y_{i}$ for $i=1,2,...,l$ be the set of vertices from $T^{'}_{t_{n_{l}}}$
colored by $m_{i}$.
Let $D_{i,j}$ be a set of vertices from $C_{i} \backslash (X \cup Y_{1} \cup ... \cup Y_{q})$ colored by color $j$. Denote by $N^{D_{i,j}}_{\rho} \subseteq D_{i,j}$ the subset of $D_{i,j}$ consisting of vertices of $D_{i,j}$ adjacent from $\rho$ if $\rho$ is before all vertices of $D_{i,j}$ in the $m$-sequence and adjacent to $\rho$ otherwise (the ordering in the $m$-sequence is induced by an ordering of sets $C_{i}$ and a transitive ordering within transitive parts of the $m$-sequence). 
Similarly, denote by $N^{D_{i,j}}_{r_{u}} \subseteq D_{i,j}$ for $u=1,2,...,q$ the subset of $D_{i,j}$ consisting of vertices of $D_{i,j}$ adjacent from $r_{u}$ if $r_{u}$ is before $D_{i,j}$ in the $m$-sequence and adjacent to $r_{u}$ otherwise. 
Denote $N^{D_{i,j}} = N^{D_{i,j}}_{r}\cap N^{D_{i,j}}_{\rho}$, where: $N^{D_{i,j}}_{r}=\bigcap_{u=1,2,...,q} N^{D_{i,j}}_{r_{u}}$. 
We run algorithm~\textit{MakeSmooth} on the given $m$-sequence, where we have: $S^{j}_{i}=N^{D_{i,j}}$, parameter $f$ satisfies: $(1-f)=\xi(1-\frac{\lambda h}{\xi})$ and get a new smooth $m$-sequence. In this new $m$-sequence the coloring is inherited from the old one. Now we rerun algorithm~\textit{FindStrong-M-Sequence-Main} with updated parameters $c,\lambda. \xi$: $c \rightarrow \frac{c\xi}{2}(1-\frac{\lambda h}{\xi})$, $\lambda \rightarrow \frac{4\lambda k(h+1)}{\xi^{2}(1-\frac{\lambda h}{\xi})^{2}}$, $\xi \rightarrow \frac{\xi}{2}(1-\frac{\lambda h}{\xi})$, $\sigma \rightarrow \sigma \setminus \{\Sigma^{*}\}$. 
\end{itemize}
\end{algorithm}

\section{Analysis of the algorithms}

In this section we formally prove correctness of the algorithm that colors $H$-free tournaments, where $H$ is a constellation. As a corollary we prove Theorem~\ref{constelcolortheorem}.

We start with some introductory observations:

\begin{theorem}
\label{densitytheorem}
Let $T$ be a tournament.
Assume that for two disjoint subsets $X,Y \subseteq V(T)$ the following holds: $d(X,Y) \geq 1 - \lambda$ for some $\lambda < 1$. Assume that $X_{1} \subseteq X$, $Y_{1} \subseteq Y$, $X_{1} \geq c_{1} |X|$, $Y_{1} \geq c_{2} |Y|$ for some $0  < c_{1},c_{2} < 1$. Then $d(X_{1},Y_{1}) \geq 1 - \frac{\lambda}{c_{1}c_{2}}$.
\end{theorem}

\Proof
Let $e_{Y,X}$ be the number of directed edges from $Y$ to $X$ and let $e_{Y_{1},X_{1}}$ be the number of directed edges from $Y_{1}$ to $X_{1}$.
We have: $e_{Y,X}=(1-d(X,Y))|X||Y| \leq \lambda|X||Y|$, since $d(X,Y) \geq 1 - \lambda$. Similarly: $e_{Y_{1},X_{1}}=(1-d(X_{1},Y_{1}))|X_{1}||Y_{1}|$.
Assume by contradiction that $d(X_{1},Y_{1}) < 1 - \frac{\lambda}{c_{1}c_{2}}$. Then, since  $X_{1} \geq c_{1} |X|$, $Y_{1} \geq c_{2} |Y|$, we have:  $e_{Y_{1},X_{1}} > \lambda |X||Y|$. Since $e_{Y,X} \geq e_{Y_{1},X_{1}}$, we get: $e_{Y,X} > \lambda |X||Y|$, contradiction.
\bbox

\begin{theorem}
\label{coloringlemma}
Assume that every subtournament $T_{S}$ of a tournament $T$ contains a transitive subtournament of order at least $|T_{S}|^{\epsilon}$ for some $\epsilon > 0$. Then $\chi(T) \leq n^{1-\epsilon}\log(n)$.
Besides if in every subtournament $T_{S}$ of $T$ one may find a transitive subtournament of order at least $|T_{S}|^{\epsilon}$ in time $O(h(|T_{S}|))$ for some nondecreasing function $h$, then the proper coloring of $T$ using at most $n^{1-\epsilon}\log(n)$ colors may be constructed in time 
$O(n^{1-\epsilon}\log(n) h(n) + n^{2}\log(n))$. 
\end{theorem}

\Proof
In the preprocessing phase we sort each adjacency list. This requires $O(n^{2} \log(n))$ time.
Find a transitive subtournament $T_{1}$ of $T$ with $|T_{1}| \geq (\frac{n}{2})^{\epsilon}$ and delete it from $T$.  To perform a deletion we first sort the vertices of the found tournament and this can be done
in $O(n\log(n))$ time. Then we get rid of all the adjacency lists that are related to the vertices from the found tournament. This can be done in $O(n\log(n))$ time simply by going through each adjacency list
and performing a binary search in the sorted sequence of the vertices from the transitive subtournament.
Finally we delete vertices of the transitive subtournament from all remaining adjacency lists and this can be done in $O(n|T_{1}|\log(n))$ time.
We keep finding transitive subtournaments of order at least $(\frac{n}{2})^{\epsilon}$ as long there are at least $\frac{n}{2}$ vertices in the tournament. The total time spent for running this subprocedure is: $O(h(n)n^{1-\epsilon} + n\log(n) + n|T_{1}|\log(n))$. 
When we reach the state with less than $\frac{n}{2}$ vertices remaining, we have found $O(n^{1-\epsilon})$ transitive subtournaments: $T_{1},T_{2},...$. We then apply the same subprocedure on the remaining graph of less than $\frac{n}{2}$ vertices. We stop when there are no vertices left and by that time we have partitioned tournament $T$ into transitive subtournaments.
If we denote by $H(n)$ the number of the transitive subtournaments found then we have the following simple recurrence formula: $H(n) \leq (\frac{n}{2})^{1-\epsilon} + H(\frac{n}{2})$, which immediately gives us: $H(n) \leq n^{1-\epsilon}\log(n)$. By coloring each transitive tournament with the same color and using different colors for different transitive subtournaments we get a proper coloring of $T$ that uses at most $n^{1-\epsilon}\log(n)$ colors. If we denote by $T(n)$ the total running time of the algorithm then
the above observations (and simple calculations) give us the following formula: $T(n)=O(n^{1-\epsilon}\log(n) h(n) + n^{2}\log(n))$. That completes the proof. 
\bbox

The following theorem turns out to be very important to prove the correctness of our coloring algorithm.
It also explains how a polynomial lowe bound on the size of the transitive subtournament can be obtained.

\begin{theorem}
\label{merginglemma}
Let be $T$ be a tournament and let $A_{1},T_{1} \subseteq V(T)$. Denote by $T^{A_{1}}$ a tournament induced by $A_{1}$. Assume that $(A_{1},T_{1})$ is $(c,\epsilon_{c})$-saturated, $c>0$, and $T^{A_{1}}$ is $\epsilon_{c}$-transitive, where $\epsilon_{c}=\frac{\log(1-c)}{\log(c)}$. Assume furthermore that a transitive subtournament of $T^{A_{1}}$ of size at least $|A_{1}|^{\epsilon_{c}}$ might be found in time $h(|A_{1}|)$ for some function $h$. Then $T$ is $\epsilon_{c}$-transitive and its transitive subtournament of size at least $|T|^{\epsilon_{c}}$ might be found in time $h(|A_{1}|)+t_{m}$, where $t_{m}$ is the size of the largest transitive subtournament of $T$.
\end{theorem}

\Proof
Denote $n=|T|$.
Let $T^{'}$ be a transitive subtournament of order at least $|A_{1}|^{\epsilon_{c}}$, found in $T^{A_{1}}$. Note that if we merge it with a tournament induced by $T_{1}$ then we get a transitive subtournament. It only suffices to prove now that this bigger transitive subtournament, denote it as $T^{l}$, satisfies: $|T^{l}| \geq n^{\epsilon_{c}}$. 
We have: $|T^{l}| \geq (cn)^{\epsilon_{c}} + cn^{\epsilon_{c}}$, since $|A_{1}| \geq cn$. Since $\epsilon_{c}=\frac{\log(1-c)}{\log(c)}$, we obtain: $|T^{l}| \geq n^{\epsilon_{c}}$. That completes the proof. \bbox

Now we prove correctness of the algorithm~\ref{lsequencealgol} and analyze its running time. Let us denote: $\lambda_{i}=(\frac{\lambda^{2}}{4^{k}h^{4k}})^{2^{i}h^{2i}}$ for $i=0,...,k$.

\begin{theorem}
If $|T| \geq \frac{2^{k+1}(h+1)h^{2k}}{\lambda_{k}^{hk}}$ then
algorithm~\ref{lsequencealgol} constructs a $(c,\lambda)$-$l$-sequence for $c=\frac{\lambda_{k}^{hk}}{2^{k}h^{2k}}$, where $2^{k}$ is the length of the sequence and $h$ is the size of the forbidden subtournament, and runs in polynomial time.
\end{theorem}

\Proof
Note first that $\lambda_{i} \leq \frac{\lambda_{i-1}^{2hi}}{4^{i}h^{4i}} \lambda$ for $i=1,2,...$. We call this property of the sequence $\{\lambda_{i}\}$ for $i=0,1,...$ the $\alpha$-property.
The algorithm trivially works for $k=0$ so we can assume from now on that $k>0$. Let $h=|H|$ and $n=|T|$, where $H$ is a forbidden tournament.
Note that the algorithm stops when the subroutine~\ref{lsequence2algol} is called with $h^{r}=1$ or the algorithm~\ref{lsequencealgol} itself is called with $k=0$. Note also that if the former holds then the last $h$ calls in the recursive call-tree on the path ending at that call were the calls of the subroutine~\ref{lsequence2algol}. 
But then we can take last $h$ vertices $v^{*}$ found in $h$ last calls of the subroutine~\ref{lsequence2algol} and they induce a copy of $H$ in $T$, contradiction. 
Let us explain in detail why this is the case. Take the first vertex $v^{*}$ from the sequence of $h$ consecutive ones and call it $v^{*}_{1}$.
Notice that the remaining $h-1$ consecutive calls will operate on the set of sets $\{N_{j}(v_{1}^{*})\}$. 
The remaining $h-1$ vertices $v^{*}$ induce a copy of $H \backslash \{v_{1}^{*}\}$ in such a way that in the embedding each vertex of $H \backslash \{v_{1}^{*}\}$
resides in the different set $N_{j}(v_{1}^{*})$. But then we can take this embedding of $H \backslash \{v_{1}^{*}\}$, add vertex $v_{1}^{*}$ and from the definition
of the sets $N_{j}(v_{1}^{*})$ we conclude that the constructed set of $h$ vertices induces a copy of $H$.
Thus if the algorithm stops then algorithm~\ref{lsequencealgol} is recursively called with $k=0$. Note also that the algorithm must stop since in the recursive call-tree there does not exist a path of $h$ calls of the subroutine~\ref{lsequence2algol} (from what we have said so far) and whenever algorithm~\ref{lsequencealgol} is called parameter $k$ is being decreased by $1$. 
Note that, since $n \geq \frac{2^{k+1}(h+1)h^{2k}}{\lambda_{k}^{hk}}$, whenever algorithm~\ref{lsequencealgol} is recursively called, the $H$-free tournament it operates on is of size 
at least  $2(h+1)$.
To see this, consider one call of the algorithm~\ref{lsequencealgol}, and let $i$ be its second parameter.
Note first that after one call of algorithm~\ref{lsequencealgol} and at most $h$ consecutive calls of subroutine~\ref{lsequence2algol} the $H$-free tournament $T^{2}$ which is the last parameter of the next call of algorithm~\ref{lsequencealgol} is of size at least $(\frac{|T^{1}|}{h+1}-1)\lambda_{i}^{h}\frac{1}{h-1}=|T^{1}|(\frac{1}{h+1}-\frac{1}{|T^{1}|})\frac{\lambda_{i}^{h}}{h-1}$,
where $T^{1}$ is the last parameter of the previous call of algorithm~\ref{lsequencealgol}.
Finally, note that on the path of the tree of recursive calls there are at most $k$ consecutive calls of algorithm~\ref{lsequencealgol} and that $\lambda_{k} \leq \lambda_{i}$ for $i=0,1,...,k-1$.
Assume now that in the subroutine~\ref{lsequence2algol} we reached the state when the two sets $W,S_{t_{j^{*}}}$ were found (see: description of the algorithm from the previous section). Assume without loss of generality that $d(W,S_{t_{j^{*}}}) \geq 1-\lambda_{i}$. 
Denote by $n_{0}$ the size of the tournament $T^{r}$ which is the parameter of the last call of algorithm~\ref{lsequencealgol} preceding in the call-tree the construction of $W$ and $S_{t_{j^{*}}}$. 
Note that inductively sequences $(A_{1},...,A_{2^{k-1}})$ and $(A^{'}_{1},...,A^{'}_{2^{k-1}})$  are both  $(c(H,k-1,\lambda),\lambda)$-$l$-sequences. Note also that we have: $|W|,|S_{t_{j^{*}}}| \geq s$, where $s=\lfloor\frac{n_{0}}{h+1} \rfloor \lambda_{k}^{i} \frac{1}{h-1}$ (this comes from the previous observation that between two consecutive runs of algorithm~\ref{lsequencealgol} we have at most $h$ recursive runs of the subroutine~\ref{lsequence2algol}).
We have: $s \geq (\frac{n_{0}}{h+1}-1)\lambda_{i}^{h} \frac{1}{h-1}$, thus $s \geq n_{0}(\frac{1}{h+1}-\frac{1}{n_{0}})\lambda_{i}^{h} \frac{1}{h-1}$.
Since at any point of the execution of the algorithm an $H$-free tournament we are dealing with has size at least $2(h+1)$, we can conclude that $|W|,|S_{t_{j^{*}}}| \geq f n_{0}$, where $f=\frac{\lambda_{i}^{h}}{2(h^{2}-1)}$. 
Then, since $\lambda_{i} \leq \lambda c^{2}(H,i-1,\lambda)$ (which follows from the $\alpha$-property) and $c(H,i,\lambda)\leq fc(H,i-1,\lambda)$, using Theorem~\ref{densitytheorem}, we can deduce that the sequence $(A_{1},...,A_{2^{k-1}},A^{'}_{1},...,A^{'}_{2^{k-1}})$ is a $(c(H,k,\lambda),\lambda)$-$l$-sequence.

To prove that algorithm~\ref{lsequencealgol} runs in polynomial time, note that time spent by the algorithm between two its recursive consecutive runs on the path of recursive calls is polynomial. Therefore, if $T(k)$ denotes time spent by the algorithm to find a $l$-sequence of length $2^{k}$, then we have:
$T(k) \leq \textit{poly}(n) + 2T(k-1)$. Since $k$ is a constant, $T(k)$ is clearly polynomial in $n$.
\bbox

Now we prove correctness of the algorithm~\ref{cliquealgorithm}.

\begin{theorem}
Algorithm~\ref{cliquealgorithm} computes a clique $\{v_{1},...,v_{k}\}$.
\end{theorem}

\Proof
Assume first that at each stage of the algorithm sets $N^{v}_{i}$ are nonempty. We will prove it later.
Note that from Theorem~\ref{densitytheorem} we know that $|W_{i}| \leq \frac{|V_{1}|}{2k}$. Thus we have $|W_{2} \cup ... \cup W_{k}| \leq |W_{2}| \cup ... \cup |W_{k}| \leq k \frac{|V_{1}|}{2k} \leq \frac{|V_{1}|}{2}$. Thus a set $V_{1} \backslash (W_{2} \cup ... \cup W_{k})$ is nonempty so we can always find $v_{1}$. It is obvious that if we combine the clique found in the next call of the algorithm~\ref{cliquealgorithm} with vertex $v_{1}$ then we get a clique. Note also that since $|N^{v}_{i}| \geq (1-2k\lambda)|V_{i}|$, by Theorem~\ref{densitytheorem}, we have: $d(V^{'}_{i},V^{'}_{j}) \geq 1 - \frac{\lambda}{(1-2k\lambda)^{2}}$ for $2 \leq i < j \leq k$, so we update the parameters of the algorithm correctly.
Notice that the first run of the algorithm~\ref{cliquealgorithm} is for $\lambda \leq \frac{1}{k3^{2k+1}}$ and that altogether there are exactly $k$ calls of the algorithm, where in each call we update: $\lambda \rightarrow \frac{\lambda}{(1-2k\lambda)^{2}}$. Thus each time the algorithm~\ref{cliquealgorithm} is called we have: $\lambda \leq \frac{1}{3k}$. Therefore in particular, whenever sets $N^{v}_{i}$ are calculated they are always nonempty. That completes the proof of the correctness of the algorithm~\ref{cliquealgorithm}.  
\bbox

Now we prove correctness of the algorithm~\ref{msequencealgol} and analyze its running time.

\begin{theorem}
Algorithm~\ref{msequencealgol} computes a $(\min(c,(\frac{c}{2})^{\epsilon}),\Lambda,\epsilon)$-$m$-sequence which is $(\log(cn)-2)$-big and runs in time $O(h(A^{m})n^{1-\epsilon}) + 
\textit{poly}(n)$, where $\textit{poly}(n)$ is a polynomial factor.
\end{theorem}

\Proof
Denote $n=|T|$.
Denote $A^{p}_{2i}= T^{i}_{1} \cup ... \cup T^{i}_{r_{i}}$ for $i=1,2,...,k$. Note that $|A^{p}_{2i}| \geq \frac{|A_{2i}|}{2}$ for $i=1,2,...,k$. Denote $A^{'}_{2i+1}=A_{2i+1}$ and $A^{'}_{2i}=A^{p}_{2i}$ for $i=1,2,...,k$. Then, using Theorem~\ref{densitytheorem}, we can deduce that $d(A^{'}_{i},A^{'}_{j}) \geq 1 - 4 \lambda$ for $1 \leq i < j \leq 2k+1$.
Indeed, $|A^{'}_{i}| \geq \frac{|A_{i}|}{2}$ and $|A^{'}_{j}| \geq \frac{|A_{j}|}{2}$ and furthermore:
$d(A_{i},A_{j}) \geq 1 - \lambda$.
Now take some $A^{'}_{i}$ and $A^{'}_{j}$ for $1 \leq i < j \leq 2k+1$ and corresponding sets $V_{i}$ and $V_{j}$ in $G$. Note that since $d(A^{'}_{i},A^{'}_{j}) \geq 1 - 4 \lambda$, using Theorem~\ref{densitytheorem}, we can conclude that the number of edges going between $V_{i}$ and $V_{j}$ is at least $(1-\lambda_{0})|V_{i}||V_{j}|$, where $\lambda_{0}=\frac{1}{(2k+1)3^{4k+3}}$. 
Let us explain in detail why this is the case. Assume otherwise. Then there are at least 
$\lambda_{0}|V_{i}||V_{j}|$ pairs of elements $(x,y)$ from $V_{i}$ and $V_{j}$ that are not adjacent
in $G$ (notice that each $x$ and $y$ is a subset). Each element from $x$ has the same size and each element $y$ has the same size (even though the size of $x$ does not have to be the same as the size of $y$). Denote the size of each element of $V_{i}$ by $s_{1}$ and the size of each element of $V_{j}$ by $s_{2}$. 
Then we can conclude that there are more than $\lambda_{0}|V_{i}||V_{j}|s_{1}s_{2}\Lambda$
edges going from $A^{'}_{j}$ to $A^{'}_{i}$ (since by definition of the nonedge of $G$, each nonedge $(x,y)$ introduces at least $\Lambda |x| |y|$ edges from $A^{'}_{j}$ to $A^{'}_{i}$).
Notice that the size of $A^{'}_{i}$ is $|V_{1}|s_{1}$ and the size of $A^{'}_{j}$ is $|V_{2}|s_{2}$.
Thus there are more than $\lambda_{0}\Lambda|A^{'}_{i}||A^{'}_{j}|$ edges going from $A^{'}_{j}$ to $A^{'}_{i}$.
On the other hand, since $d(A^{'}_{i},A^{'}_{j}) \geq 1 - 4\lambda$, we know that the number of edges
going from $A^{'}_{j}$ to $A^{'}_{i}$ is at most $4 \lambda |A^{'}_{i}||A^{'}_{j}| \leq \lambda_{0} \Lambda |A^{'}_{i}||A^{'}_{j}|$, where the last inequality follows from the fact that (by assumptions of the theorem) $\lambda \leq \lambda_{0}\frac{\Lambda}{4}$. We get a contradiction.
Thus indeed the number of edges going between $V_{i}$ and $V_{j}$ is at least $(1-\lambda_{0})|V_{i}||V_{j}|$.
But then we see that all conditions necessary to run algorithm~\ref{cliquealgorithm} are satisfied. 
The parameter $k$ in the statement of the algorithm~\ref{cliquealgorithm} correspond to $2k+1$ in our setting since our $m$-sequence is of length $2k+1$. Thus $\lambda_{0}$ from our setting corresponds to the upper bound on $\lambda$ from the statement of algorithm~\ref{cliquealgorithm}.

Note also that a clique found by this algorithm in $G$ corresponds to the $(c^{'},\Lambda,\epsilon)$-$m$-sequence for $c^{'}=\min(c,(\frac{c}{2})^{\epsilon})$. This comes from the fact that each $T^{i}_{j}$ satisfies $|T^{i}_{j}| \geq (\frac{c}{2})^{\epsilon} n^{\epsilon}$.
This $m$-sequence is $(\log(cn)-2)$-big since each extracted $T^{i}_{j}$ satisfied: $|T^{i}_{j}| \geq \log(cn)-2$. This is the case since tournaments $T^{i}_{j}$ are extracted from tournaments of size at least $s \geq \frac{cn}{2}$ and by classic Ramsey argument, each such tournament has a transitive subtournament of order at least $\log(s) - 1$.

Ley us analyze now the running time of the algorithm. Note first that if $W$ is an $N$-vertex tournament then its transitive subtournament of size at least $\log(N)-1$ may be found as follows:
Take an arbitrary vertex $v_{1} \in V(W)$. Let $N^{-}_{v_{1}}$ be the set of its inneighbors in $W$ and let $N^{+}_{v_{1}}$ be the set of its outneighbors in $W$. If $N^{-}_{v_{1}} \geq \frac{N-1}{2}$ then let $W_{1}=W|N^{-}_{v_{1}}$, otherwise let $W_{1} = W|N^{+}_{v_{1}}$. Now consider a tournament $W_{1}$ and repeat the procedure by taking an arbitrary vertex $v_{2} \in V(W_{1})$ and considering sets of its inneighbors and outneighbors in $W_{1}$, etc. Using this procedure we get the sequence of vertices: $v_{1},...,v_{r}$ for some $r$. It is easy to see that $r \geq \log(N)-1$ and that $\{v_{1},...,v_{r}\}$ is a transitive subset. Trivial implementation of this algorithm clearly runs in $\textit{poly}(N)$ time.   
We use the procedure we have just described to find transitive subtournament of size at least $\log(\frac{cn}{2})-1$ in algorithm~\ref{msequencealgol}. 
Note that the only possibly nonpolynomial part of the running time corresponds to extracting transitive subtournaments with the use of procedure $h$. Fix some 
$A_{i}$. Each transitive subset extracted from $A_{i}$ is of size at least $(\frac{|A_{i}|}{2})^{\epsilon}$. Thus the number of extracted transitive subsets from any given $A_{i}$ is $O(|A_{i}|^{1-\epsilon})=O(n^{1-\epsilon})$. Extracting a transitive subset requires time $O(h(|A_{i}|))=O(h(A^{m}))$. Finally note that we have a fixed number of sets $A_{i}$. That completes the analysis of the running time of the algorithm.
\bbox 

Now we prove correctness of the algorithm~\ref{smoothingalgorithm}.

\begin{theorem}
Algorithm~\ref{smoothingalgorithm} computes a smooth $(c^{'},\lambda^{'},\epsilon)$-$m$-sequence, a smooth $(c^{'},\lambda^{'},\epsilon)$-$t$-sequence or a smooth $(c^{'},\lambda^{'})$-$l$-sequence $(C^{'}_{1},...,C^{'}_{k})$ respectively, where $c^{'}=\frac{c}{2}(1-f)$, $\lambda^{'}=\frac{4\lambda L}{(1-f)^{2}}$, $L=s_{1}+...+s_{k}$ and $C^{'}_{i} \subseteq \bigcup_{j=1,2,...,s_{i}} S^{j}_{i}$ for $i=1,2,...,k$.  Besides we have: $|C^{'}_{i} \cap S^{j}_{i}| \geq \frac{cn}{2}(1-f)$ for $i=1,2,...,k$, $j=1,2,...,s_{i}$.
\end{theorem}

\Proof
We assume without loss of generality that a $(c,\lambda,\epsilon)$-$m$-sequence is given in the input.
Let $T$ be a tournament with the $(c,\lambda,\epsilon)$-$m$-sequence and denote $n=|T|$.
Note first that from Theorem~\ref{densitytheorem} we know that $d(S^{t_{1}}_{i},S^{t_{2}}_{j}) \geq 1 - \frac{\lambda}{(1-f)^{2}}$. Using Theorem~\ref{densitytheorem} again we can conclude that $|S^{t_{1}}_{i} \backslash C^{j,t_{2}}_{i,t_{1}}| \leq \frac{|S^{t_{1}}_{i}|}{2L}$. 
To see that assume without loss of generality that $i<j$ (for $i>j$ the analysis is exactly the same). Note first that $d(S^{t_{1}}_{i},S^{t_{2}}_{j}) \geq 1 -\frac{\lambda}{ (1-f)^{2}}$.
Thus the number of directed edges from $S^{t_{2}}_{j}$ to $S^{t_{1}}_{i}$ is at most $\frac{\lambda}{(1-f)^{2}}|S^{t_{1}}_{i}||S^{t_{2}}_{j}|$.
However if  $|S^{t_{1}}_{i} \backslash C^{j,t_{2}}_{i,t_{1}}| > \frac{|S^{t_{1}}_{i}|}{2L}$ then the number of vertices of $S^{t_{1}}_{i}$ that are adjacent to
less than $(1-\frac{2L\lambda}{(1-f)^{2}})|S^{t_{2}}_{j}|$ vertices of $S^{t_{2}}_{j}$ is more than $\frac{|S^{t_{1}}_{i}|}{2L}$ which clearly indicates that the number of directed edges from $S^{t_{2}}_{j}$ to $S^{t_{1}}_{i}$ is more than $\frac{2L\lambda}{(1-f)^{2}}|S^{t_{2}}_{j}||\frac{S^{t_{1}}_{i}|}{2L}$. This is a contradiction according to what we have noted before.

Therefore we have: $|C^{i,t_{1}}| = |S^{t_{1}}_{i} \backslash \bigcup_{j \neq i ,t_{2}=1,...,s_{j}} (S^{t_{1}}_{i} \backslash C^{j,t_{2}}_{i,t_{1}})| \geq |S^{t_{1}}_{i}|-L \frac{|S^{t_{1}}_{i}|}{2L} \geq \frac{|S^{t_{1}}_{i}|}{2}$ (notice that the number of terms in the sum from the last sequence of inequalities is at most $L$).
Now take a vertex $v \in C^{i,t_{1}}$ and take some $j>i$, $t_{2} \in \{1,2,...,s_{j}\}$. Note that there are at least $(1-\frac{2L\lambda}{(1-f)^{2}})|S^{t_{2}}_{j}|$ vertices in $|S^{t_{2}}_{j}|$ adjacent from $v$. Since $|C^{j,t_{2}}| \geq \frac{|S^{t_{2}}_{j}|}{2}$, we can conclude that $d(\{v\},C^{j,t_{2}}) \geq 1 - \frac{\frac{2L\lambda}{(1-f)^{2}})|S^{t_{2}}_{j}|}{\frac{|S^{t_{2}}_{j}|}{2}}$.
Thus $d(\{v\},C^{j,t_{2}}) \geq 1 - \frac{4\lambda L}{(1-f)^{2}}$.
Similar analysis can be done for $j<i$, $t_{2} \in \{1,2,...,j\}$. 
Now note that $|C^{'}_{i} \cap S^{j}_{i}| \geq \frac{|S^{j}_{i}|}{2} \geq \frac{(1-f)cn}{2}$.
That completes the proof.
\bbox

Now we prove correctness of the algorithm~\ref{strongstructure}.

\begin{theorem}
Algorithm~\ref{strongstructure} computes an $\mathcal{I}$-strong $(\hat{c},\hat{\lambda},\epsilon)$-$m$-sequence of length $k{'}$ which is $\frac{1}{2^{2^{4h+3}}}M$-big, where $\hat{c}=\frac{1}{2^{2^{4h+3}}} c$, $\hat{\lambda}=2^{2^{5h}}\lambda$, $\mathcal{I}=\{h+1,2h+1,...,h^{2}+1\}$ and $k^{'}=2h^{2}+4h+1$, assuming that $\lambda \leq \frac{1}{2^{2^{5h+6}}}$ and $M \geq 2(h+2) \cdot 2^{2^{8h+2}}$.
\end{theorem}

\Proof
Note that from the definition of a graph $G$ and the way it is updated it is clear that when the algorithm stops it returns an $\mathcal{I}$-strong $m$-sequence. Indeed, every clique $\{C_{i_{1}},C_{i_{2}},...,C_{i_{h}}\}$ in $G$ satisfies: $d(C_{i_{j_{1}}},C_{i_{j_{2}}})=1$ for $1 \leq j_{1} < j_{2} \leq h$.
Now note that state $0$ may be achieved throughout the execution of the algorithm at most $\frac{|V(G)|^{2}}{2}(1-\frac{1}{h-1})+1$ times, where $|V(G)|=2^{h+1}$. This is true since when state $0$ is achieved a new edge is added to $G$. Graph $G$ cannot have more than $\frac{|V(G)|^{2}}{2}(1-\frac{1}{h-1})+1$ edges since, by Turan's Theorem, if $G$ has $\frac{|V(G)|^{2}}{2}(1-\frac{1}{h-1})+1$ edges then it has a clique of size $h$. 
Note also that state $1$ may be achieved at most $h-1$ times in a row since otherwise we can merge stars found at each of the consecutive stages when state $1$ was achieved to reconstruct $H$. That contradicts the fact that $T$ is $H$-free.
We can conclude that the algorithm stops.
It only suffices to show that all its parameters are correctly updated.
This comes directly from algorithm~\ref{smoothingalgorithm} that was analyzed before, Theorem~\ref{densitytheorem} and the fact that at every stage of the algorithm each transitive set of the $m$-sequence has at least $2(h+2)$ vertices (thus every expression of the form $\lfloor\frac{|T|}{h+1}\rfloor$, where $T$ is a transitive part of the $m$-sequence, may be bounded from below by $\frac{|T|}{2(h+1)}$). We call this last property an $\Omega$-property and will prove it later. 
Knowing that, we are ready to analyze in more detail the updates when states: 0 and 1 are reached.
To see that whenever state $0$ is reached all parameters are correctly updated, notice that while replacing elements of the $m$-sequence we decrease the size of each set of the sequence at most $\frac{\xi}{h}$ times. 
Now assume that state $1$ was reached.
Note that each $D_{i,j}$ satisfies: $|D_{i,j}| \geq \xi |C_{i}|$.
Since the $m$-sequence is smooth,by Theorem~\ref{densitytheorem}, we have: $d(\rho,D_{i,j}) \geq 1 - \frac{\lambda}{\xi}$. Thus we have $|N_{\rho}^{D_{i,j}}| \geq c\xi(1-\frac{\lambda}{\xi})$. Similarly: $|N_{r_{i}}^{D}| \geq c\xi(1-\frac{\lambda}{\xi})$ for $i=1,...,q$. Therefore we have: $|N^{D_{i,j}}| \geq c\xi (1-\frac{\lambda h}{\xi})$. But then we can run algorithm~\ref{smoothingalgorithm} with $(1-f)=\xi(1-\frac{\lambda h}{\xi})$. That observations enables us to finish the analysis of the parameters' updates when
state $1$ is reached.

It remains to prove an $\Omega$-property.
The fact that at every stage of the algorithm each transitive set of the $m$-sequence has at least $2(h+2)$ vertices is implied by our next remark. One can notice that under our choice of the initial values of parameters $c,\lambda$ we have: $\frac{\lambda h}{\xi} \leq \frac{1}{2}$ and $\xi \geq \frac{1}{2^{2h-1}(h+2)}$ at every stage of the algorithm. 
Note that when state $0$ is reached and we update the parameters, we have: $\xi=\frac{1}{2(h+2)}$. On the other hand, since  $\frac{\lambda h}{\xi} \leq \frac{1}{2}$, when stage $1$ is reached and we update the parameters, we have: $\xi_{new} \geq \frac{\xi_{old}}{4}$, where $\xi_{new}$ is the value of $\xi$ after the update and $\xi_{old}$ is the one before the update. Thus, since state $1$ may be achieved at most $h-1$ times in a row, we get: $\xi \geq \frac{1}{2^{2h-1}(h+2)}$ at every stage of the execution of the algorithm.
Now, from what we have said so far, we can conclude that whenever state $0$ is achieved we have: $\lambda_{new} \leq 4h^{2}k(h+2)^{2}2^{4h-2} \lambda_{old}$ and whenever state $1$ is achieved we have: $\lambda_{new} \leq 16k(h+1)(h+2)^{2}2^{4h-2} \lambda_{old}$, where: $\lambda_{new}$ is the value of $\lambda$ after the update and $\lambda_{old}$ is the one before the update.
Thus at every stage of the algorithm we also have: $\lambda \leq \frac{1}{2^{2h}(h+2)} \lambda_{init}$, where $\lambda_{init}$ is value of the parameter $\lambda$ at the very beginning of the algorithm.
Whenever state $0$ is achieved we also have: $c_{new} \geq \frac{c_{old}}{2^{2h}h(h+2)}$ and whenever state $1$ is achieved we have: $c_{new}\geq \frac{c_{old}}{2^{2h+1}(h+2)}$ ,where: $c_{new}$ is the value of $\lambda$ after the update and $c_{old}$ is the one before the update.
Now note that $k \leq 2^{2h+3}$. Notice that from what we have said before we know that state $0$ is achieved at most $2^{2h+2}$ times during execution of the algorithm~\ref{strongstructure} and state $1$ is achieved at most $(h-1)2^{2h+2}$ times.
All these observations and some simple calculations imply that at every stage of the algorithm each transitive set of the $m$-sequence is indeed of size at least $2(h+2)$. Therefore $\Omega$-property is satisfied. 

Thus an $\mathcal{I}$-strong $m$-sequence output by the algorithm is a $(\hat{c},\hat{\lambda},\epsilon)$-$m$-sequence which is $\hat{c}M$-big for parameters $\hat{c},\hat{\lambda}$ defined in the algorithm.
\bbox

Now we prove correctness of the algorithm~\ref{transitivesub}.

\begin{theorem}
Assume that we are given a $m$-sequence of the $H$-free tournament $T$ with $h=|H|$. Assume furthermore that this sequence is an $\mathcal{I}$-strong $(c,\lambda,\epsilon)$-$m$-sequence of length $k=2h^{2}+4h+1$, where $\lambda \leq \frac{1}{2^{25h^{2}}h}$, $n=|T| \geq \frac{2^{21h^{2}}}{c}$, $M \geq 2^{21h^{2}}$, $\epsilon=\frac{\log(1-\hat{c})}{\log(\hat{c})}$ and $\hat{c}=\frac{c}{2^{7h^{2}}}$. 
Then algorithm~\ref{transitivesub} computes a transitive subtournament of $T$ of order at least $|T|^{\epsilon}$.
\end{theorem}

\Proof
Our analysis is very similar to the one conducted in the proof of the correctness of the algorithm~\ref{strongstructure}. Note that it cannot be the case that during the execution of the algorithm all the stars of $H$ were found and $\sigma$ was empty at some point of the execution since after combining these stars one can reconstruct $H$ in $T$ which contradicts the fact that $T$ is $H$-free. Note also that after our choice of initial values of the parameter $\lambda$ during the entire execution we have: $\frac{\lambda h}{\xi} \leq \frac{1}{2}$ and at every point of the execution of the algorithm each set of the $m$-sequence under consideration is of size at least $6$ (this follows by simple but tedious calculations, similar to those presented in the analysis of algorithm~\ref{strongstructure}). Therefore now we can repeat analysis of the algorithm~\ref{strongstructure}. Thus we will not discuss details again. 
Note only that during the entire execution of the algorithm we have: $\xi \geq \frac{1}{4^{h+2}}$. This is true since whenever we update parameter $\xi$ we have: $\xi_{new} \geq \frac{\xi_{old}}{4}$, where $\xi_{new}$ is a value of $\xi$ after the update and $\xi_{old}$ is the one before the update. Similarly, whenever we update parameter $\lambda$ we have: $\lambda_{new} \leq 4^{2h+6}k \lambda_{old}$, where $\lambda_{new}$ is a value of $\lambda$ after the update and $\lambda_{old}$ is the one before the update. And finally, whenever we update parameter $c$ we have: $c_{new} \geq \frac{c_{old}}{4^{h+2}}$, where $c_{new}$ is a value of $c$ after the update and $c_{old}$ is the one after the update. Thus at every point of the execution of the algorithm we have: $c \geq \frac{c_{b}}{2^{7h^{2}}}$, where: $c_{b}$ is the value of the parameter $c$ at the very beginning of the algorithm.
The only new part that we will focus on concerns running procedure $\mathcal{P}$. Note that after running it we obtain an $(\frac{c}{2^{7h^{2}}},\epsilon)$-saturated pair. Thus it suffices to note that, because of Theorem~\ref{merginglemma}, under our choice of $\epsilon$, constructed transitive subtournament is of order at least $|T|^{\epsilon}$. 
\bbox

Let us prove now that the algorithm~\ref{constellationalgorithm} is correct.
We will also prove Theorem \ref{constelcolortheorem}.

\begin{theorem}
Let $H$ be a constellation of order $h$ and let $T$ be an $H$-free tournament.
Then the algorithm~\ref{constellationalgorithm} outputs a proper coloring of the vertices of $H$
that uses at most $|T|^{1-\frac{1}{2^{2^{50h^{2}+1}}}}\log(|T|)$ colors.
Besides the first color class it finds is a transitive tournament of order at least
$|T|^{\frac{1}{2^{2^{50h^{2}+1}}}}$.
\end{theorem}

\Proof
Correctness of the algorithm is a simple consequence of the fact that algorithms:\\
~\ref{lsequencealgol},~\ref{msequencealgol},~\ref{strongstructure} and~\ref{transitivesub} 
are correct. In particular, under our initial choice of 
parameter $\lambda$ we have: $c=\frac{1}{2^{2^{50}h^{2}}}$ during execution of 
the algorithm~\ref{transitivesub}. Thus we can take as an $\epsilon$ every positive 
value no greater than $\frac{\log(1-c)}{\log(c)}$. Note that in order to use 
algorithms:~\ref{lsequencealgol},~\ref{msequencealgol},~\ref{strongstructure} 
and~\ref{transitivesub} the conditions on the sizes of the elements of 
$m/l$-sequences given as an input need to be satisfied. One can check that all those conditions are satisfied 
whenever a tournament $T$ we proceed with is of size at least: $2^{2^{2^{10h^{2}+30h}}}$. 
And one can also easily note that for $\epsilon_{1} = \frac{1}{2^{2^{50h^{2}+1}}}$ we 
trivially have: $N^{\epsilon_{1}} < 2$ for $n < 2^{2^{2^{10h^{2}+30h}}}$. Thus every $n$-vertex 
tournament $T$ with $n=|T| < 2^{2^{2^{10h^{2}+30h}}}$ contains a transitive subtournament of 
size at least $|T|^{\epsilon_{1}}$. Therefore the EH coefficient of a constellation $H$ is at 
least  $\frac{1}{2^{2^{50h^{2}+1}}}$. 
That, because of Theorem~\ref{coloringlemma}, completes the proof of the correctness of 
the algorithm~\ref{constellationalgorithm}.
Let us analyze the running time of the algorithm. The only, possibly nonpolynomial factor comes from 
the execution of the algorithm~\ref{msequencealgol} and from the execution of the algorithm~\ref{transitivesub} 
(to be more precise: from extracting transitive subtournaments by procedures: 
$\textit{Sub}$ and $\mathcal{P}$). Thus let us take advantage of the analysis of the running time of 
the algorithm~\ref{msequencealgol} and the algorithm~\ref{transitivesub}. Denote by $T(n)$ the running 
time of the algorithm~\ref{transitivesub}. We have the following straightforward recursive formula: $T(n) \leq O(T((1-c)n)n^{1-\epsilon}) +\textit{poly}(n)$, 
where $\textit{poly}(n)$ is a polynomial factor. 
Similarly, if $S(n)$ is the running time of the algorithm~\ref{msequencealgol}, then we have:  
$S(n) \leq O(S((1-c)n)n^{1-\epsilon}) +\textit{poly}(n)$.
If we denote by $G(n)$ the running time of the algorithm~\ref{constellationalgorithm} then we have: 
$G(n) = O(n(T(n)+S(n))) + \textit{poly}(n)$.
Thus one can check that for $C>0$ large enough function $e^{C\log(n)^{2}}$ is an upper bound for 
$T(n)$, $S(n)$ and $G(n)$.
\bbox

\section{Further remarks}

In this section we discuss some applications of the techniques presented in the paper. 


\subsection{The Erd\H{o}s-Hajnal conjecture for constellations}

The presented algorithm gives a constructive proof of the theorem stating that every constellation satisfies the Erd\H{o}s-Hajnal conjecture. The theorem was first proven by \cite{choromanski2}. However that proof was not constructive. Besides, because it used the regularity lemma, it gave much weaker lower bounds on the EH coefficients of constellations. Our bound is also very small but needless to say, it is much bigger than the best lower bound that can be obtained with an approach that uses the regularity lemma. 
We prove that the EH coefficient of a constellation $H$ is at least $\frac{1}{2^{2^{50h^{2}+1}}}$, where $h=|H|$.

\subsection{The Erd\H{o}s-Hajnal conjecture for nonprime tournaments}

We start with one more useful notation.
For any tournament $H$ with vertex set $V(H)=\{v_{1},...,v_{h}\}$ and for any
tournaments $F_{1},...,F_{h}$ let $H(F_{1},...,F_{h})$ denote the tournament obtained from $H$ by replacing each $v_{i}$ with a copy of $F_{i}$, and making
a vertex of the copy of $F_{i}$ outadjacent to a vertex of a copy of $F_{j}$, $j \neq i$, if and only if $(v_{i},v_{j}) \in E(H)$. The copies of
$F_{i}$, $i=1,...,h$, are assumed to be vertex disjoint.

Let $H,F$ be tournaments satisfying the Erd\H{o}s-Hajnal conjecture with some $\epsilon(H),\epsilon(F)>0$. Let $V(H)=\{v_{1},...,v_{h}\}$.
Denote by $H(F,v_{2},...,v_{h})$ the tournament obtained from $H$ by replacing $v_{1}$ by $F$. We say that a tournament $H(F,v_{2},...,v_{h})$ was obtained from $H$ and $F$ by a \textit{substitution procedure} (we substitute $v_{1}$ with $F$). Note that if $|F|>1$ and $h>1$ then  $H(F,v_{2},...,v_{h})$ is not prime since $V(F)$ is a nontrivial homogeneous set. It was proven in \cite{alon} that tournament $H(F,v_{2},...,v_{h})$ also satisfies the conjecture with $\epsilon(H(F,v_{2},...,v_{h})) = \delta \epsilon(H)$ for every $\delta < \frac{\epsilon(F)}{\epsilon(H)+h\epsilon(F)}$. To be more precise, in \cite{alon} the analogous result for undirected graphs was proven and the proof of the directed version was not given explicitly. However a proof of the directed version is completely analogous to the one for the undirected version - cliques/stable sets are replaced by transitive subsets and induced subgraphs by subtournaments.
Let $\mathcal{F}$ be some family of tournaments for which the conjecture is known and let $\mathcal{\hat{F}}$ be the closure of $\mathcal{F}$ under taking substitutions. Then we can conclude that every member of $\mathcal{\hat{F}}$ also satisfies the conjecture. Besides, if we can construct a polynomial-size transitive subtournament for every member of $\mathcal{F}$ then we can also construct an algorithm that can do the same for every member of $\mathcal{\hat{F}}$. It is so since the proof that substitutions preserve the Erd\H{o}s-Hajnal property, as presented in \cite{alon}, is constructive. 
Thus as a corollary of the algorithm presented in the previous section we obtain algorithms for coloring $H$-free tournaments with $O(n^{1-\epsilon}\log(n))$ colors, where $H$ is taken from the closure $\mathcal{\hat{C}}$ of the family of constellations $\mathcal{C}$. In fact, techniques used in the algorithm from the previous section may also be used for some tournaments that are not constellations (see: the next subsection) thus we obtain coloring algorithms for even larger classes of tournaments.
We should note here that a straightforward algorithmic version of the proof that the substitution procedure preserves the Erd\H{o}s-Hajnal conjecture has a sub-exponential running time $e^{\Omega(n^{k})}$ for some constant $0<k<1$) since it needs to examine all subsets of size $\lceil n^{\delta} \rceil$ of the $n$-element set (see: \cite{alon}, pages: 4-5). Therefore whenever we use an algorithmic version of the substitution procedure we get a coloring algorithm that runs in the sub-exponential time. Since the algorithm presented in the previous section required only quasi-polynomial time, the question arises whether it is possible to get an algorithmic version of the proof of the substitution procedure that also requires only quasi-polynomial time. It seems that the method used in the proof proposed in \cite{alon} cannot be easily modified to improve the running time. However a completely different proof may potentially have this property. 
Interestingly, the algorithm proposed in the previous section to color $H$-free tournaments, where $H$ is a constellation, uses different techniques from those that were used in \cite{alon} to prove the mentioned property of the substitution procedure and that enabled us to obtain quasi-polynomial running time. An open question is whether this running time may be improved to polynomial.

\subsection{The Erd\H{o}s-Hajnal conjecture for small tournaments}

It turns out that many tournaments may be obtained from constellations by the substitution procedure. In particular, this is true for all tournaments on $5$ vertices expect for the tournament $C_{5}$ (see: Theorem~\ref{smalltourtheorem}). Thus, according to what we have said before, for all tournaments $H$ on at most $5$ vertices except the tournament $C_{5}$ we get an algorithm running on an arbitrary $H$-free tournament, finding its polynomial-size subtournament and coloring requiring only $O(n^{1-\epsilon}\log(n))$ colors. At the same time we get a constructive proof of the Erd\H{o}s-Hajnal conjecture for those tournaments.  
In fact we can say even more. It is true (though we will not show it here) that a similar method that was used in the algorithm presented in this paper may be used along with the methods presented in \cite{bcc} to give a quasi-polynomial time algorithm that colors every $n$-vertex $C_{5}$-free tournament with $O(n^{1-\epsilon(C_{5})}\log(n))$, where $\epsilon(C_{5})$ can be exactly calculated and given in closed-form (again, the regularity lemma is not required).

We do not present that algorithm in this paper because of length constraints.
The idea behind the proof is that we can construct an arbitrary $m$-sequence in the same algorithmic way as we did in this paper for constellations. This is true since the construction
of the $m$-sequence does not use the specific structure of the constellation. The only thing we need to know is that a tournament is defined by a forbidden pattern $H$.
When we have the $m$-sequence we try to reconstruct $C_{5}$ using one specific ordering of its vertices under which the graph of backward edges is a tree. Since the input
tournament is $C_{5}$-free we wont be able to succeed. Then we show that we either get a linear set exactly adjacent to/from the big transitive chunk (as in the constellation proof)
and that by induction immediately leads to the explicit bound on the size of the transitive subtournament or we obtain another graph of backward edges. The trick now is to show that this
other graph of backward edges also corresponds to $C_{5}$. That completes the proof. 
In general, whenever the nonconstructive proof is given, where the $m$-sequence is obtained at the very beginning and then some Pigeonhole Principle approach  is used to get
a linear set exactly adjacent to/from a big transitive chunk, our algorithmic framework may be used. Since we do not use the regularity lemma, we do not rely on the bounds provided by this 
tool to obtain lower bounds on the sizes on the elements of the $m$-sequence. That leads to the explicit lower bounds on the EH coefficients.

Let us also introduce few small tournaments that are not constellations but play important role in the research on the conjecture for small forbidden patterns. 
We have already introduced $C_{5}$ - a unique tournament on $5$ vertices for which every vertex has indegree $2$.
Let $T_{6}$ be a tournament with $V(T_{6})=\{1,2,...,6\}$ such that under ordering $(1,2,...,6)$ of its vertices the only backward edges are: $(4,1),(6,3),(6,1),(5,2)$.
Let $T_{6}^{1}$ be a tournament with $V(T_{6}^{1})=\{1,2,...,6\}$ such that under ordering $(1,2,...,6)$ of its vertices the only backward edges are:
$(4,1),(5,1),(5,2),(6,3)$.
Let $T_{6}^{2}$ be a tournament with $V(T_{6}^{1})=\{1,2,...,6\}$ such that under ordering $(1,2,...,6)$ of its vertices the only backward edges are:
$(1,3),(2,3),(2,4),(6,5)$.

We just note that, as in the proof presented in \cite{bcc}, the algorithm for $C_{5}$ uses two orderings of the vertices of $C_{5}$: ordering $(1,2,3,4,5)$, under which the set of backward edges is of the form $\{(4,1),(5,2),(5,1)\}$ (the so-called \textit{tree-ordering} since the graph of backward edges is a tree) and ordering $(4,1,3,5,2)$ (the so-called \textit{cyclic ordering}). 
Surprisingly, a very similar method may be used for tournament $T_{6}^{1}$ and tournament $T_{6}^{2}$. Note that both tournaments are prime. For tournament $T_{6}^{1}$ the two crucial orderings of vertices are: $(1,2,3,...,6)$, under which the set of backward edges is of the form $\{(4,1),(5,1),(5,2),(6,3)\}$ (so-called \textit{forest ordering}) and ordering $(2,4,1,6,3,5)$, under which the set of backward edges is of the form $\{(5,1),(1,2),(5,2),(3,4)\}$.  For a tournament $T_{6}^{2}$ the two crucial orderings of vertices are: $(1,2,3,...,6)$, under which the set of backward edges is of the form $\{(4,1),(5,2),(5,1),(6,3)\}$ (so-called \textit{forest ordering}) and ordering $(5,2,4,1,6,3)$, under which the set of backward edges is of the form $\{(3,4),(4,5),(3,5),(1,2)\}$. 
Thus if $\mathcal{F} = \mathcal{C} \bigcup \{C_{5},T_{6}^{1},T_{6}^{2}\}$ and $\mathcal{\hat{F}}$ is the closure of $\mathcal{F}$ under substitutions, then there exists a sub-exponential algorithm that finds a polynomial- size transitive subtournament of a $H$-free $n$-vertex tournament, where $H \in \mathcal{\hat{F}}$. Besides there exists a sub-exponential algorithm that colors any $n$-vertex $H$-free tournament with $O(n^{1-\epsilon}\log(n))$ colors, where $H \in \mathcal{\hat{F}}$.\\ 
Thus, using Theorem~\ref{smalltourtheorem}, we conclude that there exists a sub-exponential algorithm that colors every $n$-vertex $H$-free tournament with 
$O(n^{1-\epsilon}\log(n))$ colors, where $H$ is an arbitrary tournament on at most $5$ vertices or a tournament on $6$ vertices different than $T_{6}$. For those tournaments $H$ finding polynomial-size transitive subtournaments of $H$-free tournaments can be also done in sub-exponential time.

It can be proven (\cite{chorchud}) that: 

\begin{theorem}
\label{smalltourtheorem}
Every tournament on at most $5$ vertices is either isomorphic to $C_{5}$ or is of the form $H(F_{1},...,F_{h})$ for some constellation $H$ with $V(H)=\{v_{1},...,v_{h}\}$, $h>1$, and some constellations $F_{1},...,F_{h}$.
Every tournament on at most $6$ vertices is either isomorphic to $T_{6}$, $T_{6}^{1}$, or $T_{6}^{2}$ or is of the form $H(F_{1},...,F_{h})$ for some tournament $H$ with $V(H)=\{v_{1},...,v_{h}\}$, $h>1$ and some tournaments $F_{1},...,F_{h}$.
\end{theorem}

The proof uses a brute-force method thus we skip it.

Thus by using our techniques one can  obtain a constructive proof of the Erd\H{o}s-Hajnal conjecture for all tournaments on at most $5$ vertices and all tournaments on $6$ vertices but $T_{6}$. This result is interesting since in the undirected case there are still graphs on $5$ vertices for which the conjecture is open. Furthermore, the conjecture is still open for all undirected graphs on $6$ vertices that cannot be constructed from smaller graphs by the substitution procedure.




\end{document}